\documentclass[aps,superscriptaddress,eqsecnum,nofootinbib,preprintnumbers]{revtex4}
\usepackage{graphicx,epsfig}
\usepackage{amsmath}
\usepackage{amssymb}
\usepackage{subfigure}

\usepackage{relsize}

\usepackage{textcomp}

\begin{document}

\title{Constraining the Asymptotically Safe Cosmology: cosmic acceleration
without dark energy}

\author{Fotios K. Anagnostopoulos}
\email{fotis-anagnostopoulos@hotmail.com}
\affiliation{National and Kapodistrian University of Athens, Physics Department,
Panepistimioupoli Zografou,  15772, Athens, Greece}

\author{Spyros Basilakos}
\email{svasil@academyofathens.gr}
\affiliation{Academy of Athens, Research Center for Astronomy and
Applied Mathematics, Soranou Efesiou 4, 11527, Athens, Greece}
\affiliation{National Observatory of Athens, V. Paulou and I. Metaxa 15236, Penteli, Greece}

\author{Georgios Kofinas}
\email{gkofinas@aegean.gr}
\affiliation{Research Group of Geometry, Dynamical Systems and Cosmology,\\
Department of Information and Communication Systems Engineering,\\
University of the Aegean, Karlovassi 83200, Samos, Greece}

\author{Vasilios Zarikas}
\email{vzarikas@teiste.gr}
\affiliation{Theory Division, General Department, University of Thessaly, Greece}
\affiliation{Nazarbayev University, School of Engineering, Astana, Republic of Kazakhstan, 010000}

\begin{abstract}
A recently proposed Asymptotically Safe cosmology provides an elegant mechanism towards
understanding the nature of dark energy and its associated cosmic coincidence problem.
The underlying idea is that the accelerated expansion of the universe can occur due to
infrared quantum gravity modifications at intermediate astrophysical scales (galaxies or
galaxy clusters) which produce local anti-gravity sources. In this cosmological model no extra
unproven energy scales or fine-tuning are used. In this study the Asymptotically Safe model is
confronted with the most recent observational data from low-redshift probes, namely measurements
of the Hubble parameter, standard candles (Pantheon SnIa, Quasi-stellar objects), Baryonic Acoustic
Oscillations (BAOs) and high redshift probes (CMB shift parameters). Performing an overall
likelihood analysis we constrain the free parameters of the model and we test its performance
against the concordance model (flat $\Lambda$CDM) utilizing a large family of information criteria.
We find that the Asymptotically Safe model is statistically equivalent with $\Lambda$CDM, hence it
can be seen as a viable and very efficient cosmological alternative.

\end{abstract}

\maketitle

\section{Introduction}
\label{Introduction}

The discovery of a Higgs-like field at the CERN Large Hadron Collider \cite{ATLAS:2012oga}
provides a verification of the reality of the electroweak vacuum energy, and consequently
of the cosmological constant problem. The latter is nothing more than the simple statement,
due to Zeldovich \cite{Zeldovich:1967gd}, that the quantum vacuum energy density contributes
to the energy-momentum of gravity equations.

Furthermore, the accelerated expansion of the universe \cite{Riess:1998cb}-\cite{Ade:2015xua} makes
things even more intriguing. Dark energy (DE) can be due to a time varying cosmological constant
$\Lambda(z)$ (for review see \cite{ShapiroSola}, \cite{bassola}, \cite{Sool14}, \cite{gomez})
that is connected to the cosmic quantum vacuum energy, or it can be attributed to very
different possible causes, as e.g. in \cite{Kofinas:2011pq}. Usually, DE is connected with geometric
or other fields (for DE reviews see e.g. \cite{Li:2011sd}). In any case the existence of dark energy
is associated with another unsolved question, the cosmic coincidence problem. The latter urges for
an explanation of the recent passage from a deceleration era to the current acceleration stage.

In a recent work \cite{Kofinas:2017gfv}, a novel idea was proposed for solving naturally the dark
energy issue and its associated cosmic coincidence problem of recent acceleration. Remarkably, the
correct amount of dark energy was generated without the introduction of new fields in nature and
arbitrary energy scales or fine-tuning. The solution is based on the observation that the cosmic
acceleration can be due to infrared gravity modifications at intermediate astrophysical scales
which effectively generate local anti-gravity sources. Such gravity modifications are predicted
by the Asymptotic Safety (AS) program for quantum gravity \cite{ASreviews}, \cite{Falls:2014tra}.
AS makes possible every galaxy and/or galaxy clusters to be associated with a non-negligible
scale-dependent positive cosmological constant. According to the proposed explanation of DE in
\cite{Kofinas:2017gfv}, the overall net effect of all these homogeneously distributed local
anti-gravity sources through the matching between the local and the cosmic patches is the
observed recent acceleration. In this scenario the effective dark energy depends on the Newton's
constant, the galaxy or cluster length scales, and dimensionless order-one parameters predicted
by AS theory, without need of fine-tuning or introduction of new scales. Before the existence
of these astrophysical structures, anti-gravity effects are negligible. Thus, the recent cosmic
acceleration is not a coincidence but a result of the formation of Large Scale Structures (LSS)
which is taking place in the matter dominated era.

During the last decades, numerous cosmological models have been proposed to explain DE. Therefore,
it is necessary to test all of them with the criterion of how well they describe the accelerated
expansion of the universe. The characteristics of the cosmic history can be deducted based on the
luminosity distance of standard candles. The most known category is supernovae Type Ia (SnIa).
Fortunately, very recently the largest SnIa data set ever compiled has become publicly available,
namely the Pantheon sample \cite{Scolnic:2017caz}. There are also other standard candles such as
Gamma Ray Bursts (GRB) \cite{GRB} or HII galaxies \cite{Plionis:2011jj}. A second way to probe the
expansion evolution uses the angular diameter distance of standard rulers coming from clusters
or Cosmic Microwave Background (CMB) sound horizon detected through Baryon Acoustic Oscillations
(BAOs) \cite{Blake:2011en,Alam:2016hwk}, or using the CMB angular power spectrum
\cite{Aghanim:2018eyx}. Very recently, data from gravitational wave measurements, named as
``standard sirens'' \cite{Calabrese:2016bnu}, have been added to the previous techniques.

Other significant techniques refer to dynamical probes of the expansion history
\cite{Basilakos:2017rgc} which use measures of the growth rate of matter perturbations. Such
methods provide information concerning the recent cosmic history at low redshifts similarly
to Type Ia supernova data \cite{Lavre2018}. Also, galaxies which evolve on a time scale much
larger than their age dispersion are among the best cosmic chronometers \cite{Jimenez:2001gg}.
Notice that the direct measurements of the cosmic expansion play an important role in these kind
of studies. Indeed, several works have appeared in the literature that use $H(z)$ data in
constraining the dark energy models (e.g. \cite{Samushia:2006fx}, \cite{Denkiewicz:2014kna}).

The structure of the article is as follows: In Sec. \ref{SCAS} we present, along with a short
discussion of the Swiss cheese models, some theoretical aspects of the Asymptotically Safe scenario
and provide the basic cosmological equations of the proposal introduced in \cite{Kofinas:2017gfv}.
In Sec. \ref{Fitting} we discuss the method and the observational data sets that we use in order
to constraint the free parameters of the model and we compare the AS and the $\Lambda$CDM models.
Finally, we summarize our conclusions in Sec. \ref{Conclusions}.

\section{Asymptotically safe gravity and Cosmology}

\label{SCAS}

In this section we provide the main points of the Asymptotically Safe Swiss cheese model
according to \cite{Kofinas:2017gfv}, accompanied with an explanation of its motivation and its
nice properties that arise mainly from the underlying AS theory.

\subsection{Swiss cheese models}

The so-called Swiss cheese model was first introduced by Einstein and Strauss \cite{Einstein:1946ev}.
This is a solution that succeeds to describe both globally a homogeneous and isotropic universe and
locally a Schwarzschild black hole. The matching of the cosmological metric as the exterior solution
with the local interior spherical solution is proven to occur across a spherical boundary that
stays at a constant coordinate radius of the cosmological frame but evolves in the interior frame.
Here, the Swiss cheese analysis is adopted as the simplest, but not necessarily the most realistic
construction, in order to implement the idea of a cosmic acceleration driven by local repulsive
effects.

The setup consists of a 4-dimensional spacetime $M$ with metric $g_{\mu\nu}$ and a timelike
hypersurface $\Sigma$ which splits $M$ into two regions. The first important quantity is the
induced metric $h_{\mu\nu}=g_{\mu\nu}-n_{\mu}n_{\nu}$ of $\Sigma$, with $n^{\mu}$ the unit normal
vector to $\Sigma$ pointing inwards the two regions. The second one is the extrinsic curvature
$K_{\mu\nu}=h_{\mu}^{\kappa}h_{\nu}^{\lambda}n_{\kappa;\lambda}$, where a semicolon $;$ denotes
covariant differentiation with respect to $g_{\mu\nu}$. The requirement of continuity of spacetime
across $\Sigma$ implies continuity on $\Sigma$ of the parallel to $\Sigma$ components $h_{ij}$,
so $h_{ij}$ should be the same calculated on either side of $\Sigma$. In classical General
Relativity with a regular matter content and vanishing distributional energy-momentum tensor on
$\Sigma$, the Israel-Darmois matching conditions \cite{Israel:1966rt} imply that the sum of the
two extrinsic curvatures computed on the two sides of $\Sigma$ is zero. These are the two conditions
for a smooth matching.

A slight generalization of Einstein-Strauss model can be derived matching a homogeneous and
isotropic metric smoothly to a general static spherically symmetric metric.
The  Friedmann-Lemaitre-Robertson-Walker (FLRW) metric is of the form
\begin{equation}
ds^2=-dt^2+a^2(t)\left[\frac{dr^2}{1-\kappa r^2}+r^2\left(d\theta^2+\sin^2\!\theta\,d\varphi^2\right)
\right]\,,
\label{eq:FRW}
\end{equation}
where $a(t)$ is the scale factor and $\kappa=0, \pm 1$ characterizes the spatial curvature (in
the above equation we have set $c=1$). Here, a ``spherical'' boundary is described by a fixed
coordinate radius $r=r_{\Sigma}$ with $r_{\Sigma}$ constant, where of course this boundary
experiences the universal expansion. The interior part of the spacetime, with $r\leq r_{\Sigma}$,
consists of a static spherically symmetric metric given in Schwarzschild-like coordinates by
\begin{equation}
ds^2=-J(R)F(R)dT^2+\frac{dR^2}{F(R)}+R^2\left(d\theta^2+\sin^2\!\theta\,d\varphi^2\right)\,,
\label{generalBH}
\end{equation}
where $J,F>0$. Since the 2-dimensional sphere $(\theta,\varphi)$ is common for both metrics
(\ref{eq:FRW}), (\ref{generalBH}), the location of $\Sigma$ for the metric (\ref{generalBH})
is a function of $t$ only, i.e. $T=T_{S}(t), \,R=R_{S}(t)$, where the subscript $S$ is after Schucking
and $R_{S}$ is called Schucking radius.

The matching requirements provide the following equations for $R_{S}$,
\begin{eqnarray}
R_{S}&=&ar_{\Sigma}\label{eq:First FF2}\\
\Big(\frac{dR_{S}}{dt}\Big)^{2}&=&1\!-\!\kappa r_{\Sigma}^{2}-F(R_{S})\label{gyyu}\\
\frac{d^{2}R_{S}}{dt^{2}}&=&-\frac{F'(R_{S})}{2}\label{kevf}\\
J'(R_{S})&=&0\label{jrnf}
\end{eqnarray}
(with a prime denoting differentiation with respect to $R$), plus another two differential equations
for $T_{S}$ that are not important here. The successful matching has proved that the choice of
the matching surface $\Sigma$ was the appropriate one. Equations (\ref{gyyu}), (\ref{kevf})
provide through (\ref{eq:First FF2}) the cosmic evolution of the scale factor $a$. Of course,
in order for this evolution to represent a spatially homogeneous universe, not just a single
sphere of comoving radius $r_{\Sigma}$ should be present, but a number of such spheres are
uniformly distributed throughout the space. Each such sphere can be physically realized
by an astrophysical object, such as a galaxy (with its extended spherical halo) or a cluster
of galaxies.

It is common to assume that the matching radius $r_{\Sigma}$ is such that filling the interior solid
with energy density equal to the cosmic matter density $\rho$, the interior energy equals the
characteristic mass $M$ of the astrophysical object, i.e.
\begin{equation}
r_{\Sigma}=\Big(\frac{2G_{\!N}M}{\Omega_{m0}a_{0}^{3}H_{0}^{2}}\Big)^{\!\frac{1}{3}}\,,
\label{lews}
\end{equation}
where the matter density parameter is $\Omega_{m}=\frac{8\pi G_{\!N}\rho}{3H^{2}}=
\frac{2G_{\!N}M}{r_{\Sigma}^{3}a^{3}H^{2}}$ and $a_{0}$ can be set to 1 for $\kappa=0$ (which
is the case of our interest in the present work).
To give an estimate of $r_{\Sigma}$, setting $\Omega_{m0}=0.3$, for a typical galaxy mass
$M=10^{11}M_{\odot}$ we get $r_{\Sigma}=0.83\text{Mpc}$, while for a typical cluster mass
$M=10^{15}M_{\odot}$ it arises $r_{\Sigma}=18\text{Mpc}$ ($M_{\odot}$ is the solar mass).
A typical radius of a spiral galaxy (including its dark matter halo) is $R_{b}\approx 0.15\text{Mpc}$,
while the mean distance between galaxies is a few $\text{Mpc}$, thus $r_{\Sigma}$ is above the
galaxy size and the Schucking radii of two neighboring galaxies do not overlap, which is
interesting although the Swiss cheese is only an approximation to the real cosmos.
Similarly for the clusters, the radii have a larger range from $R_{b}\approx 0.5\text{Mpc}$ to
$R_{b}\approx 5\text{Mpc}$, and depending on the distance between the outer boundaries of two
nearby clusters, still no overlapping can occur between the Schucking regions.

The proposal of the work \cite{Kofinas:2017gfv} is that the dark energy is related to the presence
of structure, not due to averaging processes of inhomogeneities/anisotropies in the universe
\cite{inho}, but mainly due to the existence of anti-gravity sources related to these astrophysical
structures. Therefore, since these matter configurations are formed recently at relatively low
redshifts, dark energy is also generated recently, possibly explaining the coincidence as a consequence
of the structure. Various gravitational theories could be investigated in the context of this
scenario and test to what extent of the parameters the corresponding cosmologies comply with the
apparent acceleration, the relation $\rho_{0}\sim\rho_{DE,0}$ and other data.

When a constant cosmological term is added to the Schwarzschild metric, the well-known
Schwarzschild-de Sitter metric arises,
\begin{equation}
ds^2=-\Big(1-\frac{2G_{\!N} M}{R}-\frac{1}{3}\Lambda R^2\Big)dT^2
+\frac{dR^2}{1-\frac{2G_{\!N} M}{R}-\frac{1}{3}\Lambda R^2}
+R^2\left(d\theta^2+\sin^2\theta d\varphi^2\right)\, .
\label{ASFM}
\end{equation}
It is apparent that $J(R)=1$, satisfying (\ref{jrnf}). This metric can be motivated from
quantum gravity inspired or modified gravity models that justify the appearance
of this constant value $\Lambda$. Using (\ref{eq:First FF2}), (\ref{lews}), equations (\ref{gyyu}),
(\ref{kevf}) take the form
\begin{eqnarray}
H^{2}+\frac{\kappa}{a^{2}}&=&\frac{8\pi G_{\!N}}{3}\rho+\frac{\Lambda}{3}\label{wiiw}\\
\frac{\ddot{a}}{a}&=&-\frac{4\pi G_{\!N}}{3}\rho+\frac{\Lambda}{3}\label{erjq}
\end{eqnarray}
with $H=\dot{a}/a$ the Hubble parameter.
Although equations (\ref{wiiw}), (\ref{erjq}) are the standard cosmological equations of the
$\Lambda\text{CDM}$ model, here the $\Lambda$ term appearing in the black hole metric
(\ref{ASFM}) has a very distinctive meaning. First, this $\Lambda$ is of astrophysical origin,
an average of all anti-gravity sources inside the Schucking radius of a galaxy or cluster,
or it can even arise from astrophysical black holes in the centers of which the presence of a
repulsive pressure could balance the attraction of gravity avoiding the singularity. Second, it
is expected that it arises due to quantum corrections of a concrete quantum gravity theory.
Third, before the formation of all these structures, the total $\Lambda$ is zero, thus in reality
the $\Lambda$ of (\ref{wiiw}), (\ref{erjq}) is a function of the redshift $z$, which decreases at
larger $z$, possibly solving the coincidence problem, while $\Lambda\text{CDM}$ model, with
a universal constant related to the vacuum energy, cannot.

The Schwarzschild-de Sitter metric can be progressed to a quantum improved Schwarzschild-de Sitter
metric with the hope to describe more accurately the proper astrophysical object. This metric has
the form
\begin{equation}
ds^2=-\Big(1-\frac{2G_k M}{R}-\frac{1}{3}\Lambda_k R^2\Big)dT^2
+\frac{dR^2}{1-\frac{2G_k M}{R}-\frac{1}{3}\Lambda_k R^2}
+R^2\left(d\theta^2+\sin^2\theta d\varphi^2\right)\,,
\label{ASBH}
\end{equation}
where now both Newton's constant $G_k$ and cosmological constant $\Lambda_k$ are functions of a
characteristic energy scale $k$ of the system determined by the quantum gravity theory in use. In
principle this energy scale $k$ is expected to be a function of the distance from the center of the
spherical symmetry. However, we are interested to study the cosmic evolution through the Swiss cheese
approach, which means that only the value $R_{S}$ (or the corresponding energy value $k_{S}$) is
relevant since only this enters the differential equations (\ref{gyyu}), (\ref{kevf}). If $G_{k}$
is almost equal to the constant observed value $G_{\!N}$, while the quantum gravity model provides
only ultraviolet corrections to $\Lambda$, then cosmic acceleration cannot be generated due to the
negligible magnitude of $\Lambda$ at large distances of the order of Schucking radius. Therefore,
the existence of infrared (IR) corrections of $\Lambda$ are vital.

\subsection{Asymptotically Safe Gravity as solution of the problem}

Asymptotic Safety is the framework that provides naturally the IR corrections that are needed for
a successful effective dark energy model of astrophysical origin. AS can provide and justify
$G_{k},\Lambda_{k}$ as functions of the energy scale $k$. The dimensionless running
couplings $g_{k},\lambda_{k}$, defined by
$G_k=G(k)=g_{k}k^{-2}\,,\,\Lambda_k=\Lambda(k)=\lambda_{k}k^{2}$, are governed by some
Renormalization Group (RG) flow equations. The analytical expressions for
$g_{k},\lambda_{k}$ are not completely clear of ambiguities, so it is not still very accurately
known what is the real trajectory in the space of $g_{k},\lambda_{k}$ that was followed by our
universe. Therefore, it is not definitely known at which exactly point of this trajectory the
classical General Relativity regime, with a constant $G_{\!N}$ and negligible $\Lambda$, arises.

AS attracted the interest of the scientific community since at the transplanckian energy scales
($k\rightarrow \infty$) a Non-Gaussian fixed point (NGFP) exists \cite{Reuter:2009kq},
where $g_{k},\lambda_{k}$ become constant ($G$ approaches zero, while $\Lambda$ diverges) and
gravity becomes renormalisable in a non perturbative way. There is another fixed point, the
Gaussian fixed point (GFP) \cite{Reuter:2001ag} at $g=\lambda=0$, where in its linear regime, $G$ is
approximately constant and $\Lambda$ displays a running behavior proportional to $k^{4}$. Finally,
at the far infrared limit ($k \rightarrow 0$) \cite{Bonanno:2001hi}, the behavior of the RG flow
trajectories with positive $G,\Lambda$ cannot be accurately predicted since the RG running stops to
be valid as $\lambda_{k}$ reaches $1/2$, where an unbounded growth of $G$ appears together with a
vanishingly small $\Lambda$ (interestingly enough the divergence of the beta functions happens near
$k=H_{0}$), therefore the exact current value of $\Lambda$ is unknown. Since the present universe
possesses rather an anti-screening instead of a screening behavior, the far infrared limit will most
probably describe the future late-times cosmic evolution. The proposal in \cite{Kofinas:2017gfv} does
not concern the far infrared energy scales, but refers to the intermediate infrared corrections of
astrophysical structures scales. It was shown that the observed dark energy component and the recent
cosmic acceleration can be generated by such AS quantum corrections of the cosmological constant at
the galactic or cluster of galaxies scale without fine tuning or any obvious conflict with local
dynamics. The Swiss cheese cosmological model was used with interior metric the AS inspired
quantum improved Schwarzschild-de Sitter form (\ref{ASBH}) \cite{Koch:2014cqa}.

As mentioned, the energy measure $k$ has to be associated with a characteristic length scale $L$,
$k=\xi/L$, where $\xi$ is a dimensionless order-one number. The reasoning is that the closer to the
center of spherical symmetry, the larger the mean energy is expected to be. A simple option is to set
as $L$ the radial distance $R$ from the center, which is not particularly successful in our proposal
and will not be considered further. A more natural option is to set as $L$ the proper distance $D>0$
\cite{Bonanno:2001xi}, which is also the choice that leads to singularity avoidance/smoothening
\cite{Kofinas:2015sna}. For a radial curve with $dT=d\theta=d\varphi=0$ from $R_1$ towards a point
with coordinate $R$ is given by
\begin{equation}
D(R)=\int_{R_{1}}^{R}\frac{d\mathcal{R}}{\sqrt{F(\mathcal{R})}}\,.
\label{joet}
\end{equation}
Since $k$ enters (\ref{ASBH}), $F$ is not an explicit function of $R$ but it depends also on $D$,
so $F(\mathcal{R})$ means $F(\mathcal{R}, D(\mathcal{R}))$. Thus, (\ref{joet}) is an integral
equation which can be converted to the differential equation
\begin{equation}
D'(R)=\frac{1}{\sqrt{F(R)}}
\label{hteg}\,.
\end{equation}
This is a complicated equation since $F$ depends on $D(R)$, whose solution is $D(R;\sigma)$
with $\sigma$ an integration constant. In the Swiss cheese analysis, $\sigma$ is determined
from the correct amount of present dark energy. Since the matching of the interior with the
exterior metric happens at the Schucking radius, only $k_{S}=\xi/D_{S}$ enters the cosmological
evolution, where $D_{S}(R_{S})$ is the proper distance of the Schucking radius changing in time.
Based on (\ref{gyyu}), (\ref{eq:First FF2}), we get
\begin{equation}
H^{2}+\frac{\kappa}{a^{2}}=\frac{2G(R_{S})M}{r_{\Sigma}^{3}a^{3}}
+\frac{1}{3}\Lambda(R_{S})\,.
\label{yehg}
\end{equation}

From the AS studies, there are indications \cite{Bonanno:2001hi} that for $k\rightarrow 0$
an infrared fixed point exists, where the cosmological constant runs as
$\Lambda_{k}=\lambda_{\ast}^{\text{IR}}k^{2}$ ($\lambda_{\ast}^{\text{IR}}>0$). In
\cite{Kofinas:2017gfv}, it was reasonably assumed that, at the intermediate astrophysical scales, the
above IR fixed point has not yet been reached, but deviations from the exact $k^{2}$ law occur inside
the objects of interest. Since the astrophysical structures are still large compared to the
cosmological scales, the functional form of $\Lambda_{k}$ is expected to differ slightly from its IR
form $k^{2}$, so a power law $\Lambda_{k}\sim k^{b}$, with $b$ close to the value 2, is a fair
approximation of the running behavior \cite{Sool14}. In order for
the astrophysical value of $\Lambda_{k}$ to
get enhanced, it seems more reasonable that $b$ is slightly larger than 2, as is indeed verified by
the successful dark energy model. Additionally, the gravitational constant $G_{k}$ was
considered to have the constant value $G_{\!N}$ at observable macroscopic distances, in agreement
with the standard Newton's law. In total, our working assumption for the running couplings is
\begin{equation}
G_{k}=G_{\!N}\,\,\,\,\,\,\,\,,\,\,\,\,\,\,\,\,\Lambda_{k}=\gamma k^{b}\,,
\label{helg}
\end{equation}
where $\gamma>0, b$ are constant parameters. The dimension of $\gamma$ is mass to the power $2-b$
and it is convenient to be parametrized as $\gamma=\tilde{\gamma} G_{\!N}^{\frac{b}{2}-1}$ with
$\tilde{\gamma}$ a dimensionless number; a successful $\tilde{\gamma}$ is of order one, which means
that no new scale is needed and the coincidence problem might be explained without fine-tuning.
In addition, in order to have the correct amount of dark energy, the quantity $D$ at the Schucking
radius today, $D_{S0}$, which can be considered as an integration constant, as well as the whole
function $D(R)$, turn out to be of order $r_{\Sigma}$ that is naturally of the size of the
astrophysical object, so again no new scale is introduced. The new cosmological constant term is
$\frac{1}{3}\Lambda_{k}R^{2}=\frac{\xi^{b}\tilde{\gamma}}{3}\big[\frac{1}{G_{\!N}}
\big(\frac{\sqrt{G_{\!N}}}{r_{\Sigma}}\big)^{\!b}\big]\big(\frac{r_{\Sigma}}{D}\big)^{\!b}R^{2}$
and it seems quite interesting that for $b$ close to 2.1 the quantity
$\frac{1}{G_{\!N}}\big(\frac{\sqrt{G_{\!N}}}{r_{\Sigma}}\big)^{\!b}$ is very close to the order
of magnitude of the standard cosmological constant $\Lambda\simeq 4.7\times10^{-84}\text{GeV}^{2}$ of
$\Lambda$CDM. The factor $\big(\frac{r_{\Sigma}}{D}\big)^{\!b}$ contributes a small
distance-dependent deformation, which is however important for the precise behavior of the derived
cosmology. Therefore, the hard coincidence of the standard $\Lambda\sim H_{0}^{2}$, has here
been exchanged with a mild adjustment of the index $b$ close to 2.1, which turns out to satisfy
$\frac{1}{G_{\!N}}\big(\frac{\sqrt{G_{\!N}}}{r_{\Sigma}}\big)^{\!b}\sim H_{0}^{2}$.

To give a few more details, it was shown in \cite{Kofinas:2017gfv} that the Hubble evolution is
given by the system
\begin{eqnarray}
H^{2}+\frac{\kappa}{a^{2}}&=&\frac{2G_{\!N}M}{r_{\Sigma}^{3}a^{3}}
+\frac{\gamma\xi^{b}}{3D_{S}^{b}}\label{ksrg}\\
\dot{D}_{S}&=&\frac{r_{\Sigma}aH}{\sqrt{1-\frac{2G_{\!N}M}{r_{\Sigma}a}
-\frac{\gamma\xi^{b}r_{\Sigma}^{2}a^{2}}{3D_{S}^{b}}}}\,,\label{kewr}
\end{eqnarray}
where the variable $D_{S}$ is of geometrical nature, as explained above, with its own equation of
motion. Defining the positive variable
\begin{equation}
\tilde{\psi}=\frac{\gamma\xi^{b}}{3H_{0}^{2}D_{S}^{b}}\,,
\label{juer}
\end{equation}
which plays the role of dark energy with $\rho_{DE}=\frac{3H^{2}\Omega_{DE}}{8\pi G_{\!N}}
=\frac{3H_{0}^{2}}{8\pi G_{\!N}}\tilde{\psi}$ and $\tilde{\psi}_{0}=\Omega_{DE0}$, we obtain
the equations
\begin{eqnarray}
&&\frac{H^{2}}{H_{0}^{2}}=\Omega_{m0}(1\!+\!z)^{3}+\tilde{\psi}+\Omega_{\kappa 0}
(1\!+\!z)^{2}\label{uhne}\\
&&\frac{d\tilde{\psi}}{dz}=\frac{3^{\frac{1}{b}}b(G_{\!N}H_{0}^{2})^{\frac{1}{b}-\frac{1}{2}}
\tilde{\psi}^{1+\frac{1}{b}}}
{\xi\tilde{\gamma}^{\frac{1}{b}}(1\!+\!z)^{2}\sqrt{\frac{1}{r_{\Sigma}^{2}a_{0}^{2}H_{0}^{2}}
-\Omega_{m0}(1\!+\!z)-\frac{\tilde{\psi}}{(1+z)^{2}}}}\,,\label{jeth}
\end{eqnarray}
where $\Omega_{\kappa 0}=-\kappa/(a_{0}^{2}H_{0}^{2})$. To a very high accuracy (higher than
$0.02\%$ for clusters and higher than $0.00005\%$ for galaxies) the differential equation
(\ref{jeth}) is approximated by
\begin{equation}
\frac{d\tilde{\psi}}{dz}=\frac{3^{\frac{1}{b}}b}{\xi\tilde{\gamma}^{\frac{1}{b}}}
(G_{\!N}H_{0}^{2})^{\frac{1}{b}}\frac{r_{\Sigma}a_{0}}{\sqrt{G_{\!N}}}\,
\frac{\tilde{\psi}^{1+\frac{1}{b}}}{(1\!+\!z)^{2}}\,,
\label{yheb}
\end{equation}
whose integration gives the Hubble evolution
\begin{equation}
\frac{H^{2}(z)}{H_{0}^{2}}=\Omega_{m0}(1\!+\!z)^{3}
+\Big[\Omega_{DE0}^{-\frac{1}{b}}-\frac{3^{\frac{1}{b}}}{\xi\tilde{\gamma}^{\frac{1}{b}}}
(G_{\!N}H_{0}^{2})^{\frac{1}{b}}\frac{r_{\Sigma}a_{0}}{\sqrt{G_{\!N}}}\,\frac{z}{1\!+\!z}\Big]^{-b}
+\Omega_{\kappa 0}(1\!+\!z)^{2}\,,
\label{kieb}
\end{equation}
which is extremely good approximation for all recent redshifts that our model is interested
to cover. The positiveness of $\tilde{\psi}$ implies that
\begin{equation}
\xi^{b}\tilde{\gamma}>\frac{3\Omega_{DE0}}{(1\!+\!z_{max}^{-1})^{b}}G_{\!N}H_{0}^{2}
\Big(\frac{r_{\Sigma}a_{0}}{\sqrt{G_{\!N}}}\Big)^{\!b}\,,
\label{lbyp}
\end{equation}
where $z_{max}=\mathcal{O}(1)$ is a redshift such that the model should be defined in the range
$(0,z_{max})$. Equation (\ref{kieb}) will be tested against observations in Sec. \ref{Fitting}
for $\kappa=0$. If both sides of the inequality (\ref{lbyp}) are of the same order, all terms in
(\ref{kieb}) - with the exception of the spatial curvature - become equally important.
With a theoretically favored value of $b$ close to its IR value 2, e.g. with $b=2.13$ for galaxy or
$b=2.08$ for cluster, one obtains
$G_{\!N}H_{0}^{2}\big(\frac{r_{\Sigma}}{\sqrt{G_{\!N}}}\big)^{\!b}\sim 1$ and it can easily
be chosen $\tilde{\gamma}\sim 1$ satisfying (\ref{lbyp}), thus indeed no new mass scale is introduced
for the explanation of dark energy other than the astrophysical scales. Additionally, it can be seen
that the value of $D$ at the today matching surface is $D_{S0}\sim r_{\Sigma}$, while the
precise value of $D_{S0}$ depends on $\Omega_{DE0}$ and is determined in agreement with the
measured dark energy. If the initial condition $D_{S0}$, which is the proper distance of the
current matching surface, was not of the order of the size of the astrophysical structure,
then it would be just the integration constant of an extra field and would introduce another
unnatural scale. Not only $D_{S0}$, but the whole function $D_{S}(z)$ remains for all relevant
$z$ of the order of $r_{\Sigma}$.

The acceleration is very well approximated, as before for $H(z)$, by the expression
\begin{equation}
\frac{\ddot{a}}{H_{0}^{2}a}=-\frac{1}{2}\Omega_{m0}(1\!+\!z)^{3}+\Big[\Omega_{DE0}^{-\frac{1}{b}}
-\frac{3^{\frac{1}{b}}}{2\xi\tilde{\gamma}^{\frac{1}{b}}}(G_{\!N}H_{0}^{2})^{\frac{1}{b}}
\frac{r_{\Sigma}a_{0}}{\sqrt{G_{\!N}}}\frac{b\!+\!2z}{1\!+\!z}\Big]
\Big[\Omega_{DE0}^{-\frac{1}{b}}
-\frac{3^{\frac{1}{b}}}{\xi\tilde{\gamma}^{\frac{1}{b}}}(G_{\!N}H_{0}^{2})^{\frac{1}{b}}
\frac{r_{\Sigma}a_{0}}{\sqrt{G_{\!N}}}\frac{z}{1\!+\!z}\Big]^{-1-b}\,.
\label{ther}
\end{equation}
The current value of the deceleration parameter $q=-\ddot{a}/(aH^{2})$ obtains the form
\begin{equation}
q_{0}=\frac{1}{2}\Omega_{m0}-\Omega_{DE0}+\frac{3^{\frac{1}{b}}b}
{2\xi\tilde{\gamma}^{\frac{1}{b}}}\Omega_{DE0}^{1+\frac{1}{b}}
(G_{\!N}H_{0}^{2})^{\frac{1}{b}}\frac{r_{\Sigma}a_{0}}{\sqrt{G_{\!N}}}\;.
\label{tbem}
\end{equation}
For comparison with $\Lambda$CDM, we present here the corresponding expression,
\begin{equation}
\label{q0LCDM}
q_{0}^{\Lambda\text{CDM}}=-1+\frac{3}{2}\Omega_{m0}\,.
\end{equation}
It is easy to observe from the last two expressions that the current value of the deceleration
parameter $q_0$ in (\ref{tbem}) is smaller (as absolute value) than $q_0^{\Lambda \text{CDM}}$.

Setting $\ddot{a}=0$ in (\ref{ther}), it is possible to evaluate the transition redshift $z_{t}$
from deceleration to acceleration. The condition $\ddot{a}|_{0}>0$ is equivalent to
\begin{equation}
\xi^{b}\tilde{\gamma}>\frac{3b^{b}\,\Omega_{DE0}^{1+b}}
{(2\Omega_{DE0}\!-\!\Omega_{m0})^{b}}\,G_{\!N}H_{0}^{2}\Big(\frac{r_{\Sigma}a_{0}}
{\sqrt{G_{\!N}}}\Big)^{\!b}\,,
\label{iwrf}
\end{equation}
which is stronger than (\ref{lbyp}), so (\ref{lbyp}) can be forgotten. Finally, the effective
dark energy pressure and its equation-of-state parameter are given by (for more details see
\cite{Kofinas:2017gfv})
\begin{eqnarray}
&&\frac{8\pi G_{\!N}}{H_{0}^{2}}p_{DE}=
\Big[\frac{3^{\frac{1}{b}}}{\xi\tilde{\gamma}^{\frac{1}{b}}}(G_{\!N}H_{0}^{2})^{\frac{1}{b}}
\frac{r_{\Sigma}a_{0}}{\sqrt{G_{\!N}}}\frac{b\!+\!3z}{1\!+\!z}-3\Omega_{DE0}^{-\frac{1}{b}}\Big]
\Big[\Omega_{DE0}^{-\frac{1}{b}}
-\frac{3^{\frac{1}{b}}}{\xi\tilde{\gamma}^{\frac{1}{b}}}(G_{\!N}H_{0}^{2})^{\frac{1}{b}}
\frac{r_{\Sigma}a_{0}}{\sqrt{G_{\!N}}}\frac{z}{1\!+\!z}\Big]^{-1-b}\label{pmgu}\\
&&w_{DE}=\Big[\frac{3^{\frac{1}{b}-1}}{\xi\tilde{\gamma}^{\frac{1}{b}}}(G_{\!N}H_{0}^{2})^{\frac{1}{b}}
\frac{r_{\Sigma}a_{0}}{\sqrt{G_{\!N}}}\frac{b\!+\!3z}{1\!+\!z}-\Omega_{DE0}^{-\frac{1}{b}}\Big]
\Big[\Omega_{DE0}^{-\frac{1}{b}}
-\frac{3^{\frac{1}{b}}}{\xi\tilde{\gamma}^{\frac{1}{b}}}(G_{\!N}H_{0}^{2})^{\frac{1}{b}}
\frac{r_{\Sigma}a_{0}}{\sqrt{G_{\!N}}}\frac{z}{1\!+\!z}\Big]^{-1}\,.
\label{gbwp}
\end{eqnarray}
The present-day value is
\begin{equation}
w_{DE0}= -1+\frac{3^{\frac{1}{b}-1}b}
{\xi\tilde{\gamma}^{\frac{1}{b}}}\Omega_{DE0}^{\frac{1}{b}}
(G_{\!N}H_{0}^{2})^{\frac{1}{b}}\frac{r_{\Sigma}a_{0}}{\sqrt{G_{\!N}}}\,,
\label{ghkp}
\end{equation}
thus $w_{DE0}$ gets larger values than the $\Lambda$CDM value $-1$. To give an example of a
successful (in principle) model, for clusters with $b=2.08$, $\xi=9$, $\tilde{\gamma}=5$, it is
$q_{0}=-0.50$, $z_{t}=0.68$, $w_{DE0}=-0.95$ and the function $\Omega_{DE}(z)$ exhibits a decreasing
behavior in agreement with observations, while $q$ shows a passage from deceleration to acceleration
at late times.

As a final comment, we mention that a preliminary analysis has shown that there is no obvious
contradiction between the model discussed and the internal dynamics of the astrophysical object.
The potential and the force due to the varying cosmological constant term are small percentages
of the corresponding Newtonian potential and force, where the precise values depend on the
considered structure and the considered point at the boundary of the object or inside. Certainly,
the estimation of the anti-gravity effects inside the structure is an issue that deserves a more
thorough investigation. Therefore, our pursuit would be to show that dark energy is explained from
AS gravity and the knowledge of structure, without the introduction of new physics or new scales
and with no conflict with the local dynamics of the object.

\section{Fitting the AS Swiss cheese cosmology to the observational data}
\label{Fitting}

In this section we present the statistical methods and the cosmological data that we use in
order to put constraints on the AS Swiss cheese model described in the previous section.
Specifically, we employ direct measurements of the Hubble parameter, the standard candles
(Pantheon SnIa and Quasi-Stellar Objects) and Baryonic Acoustic Oscillations (BAOs) together
with the CMB shift parameters based on Planck 2015.

Considering a spatially flat FLRW metric, we can rewrite the Hubble parameter of the AS
cosmology (\ref{kieb}) as follows
\begin{equation}
\label{Hubble_rate_dan}
E^{2}(z,\phi^{\nu})=\Omega_{m0}(1\!+\!z)^{3}
+\Big(\Omega_{DE0}^{-\frac{1}{b}}+A\frac{z}{1\!+\!z}\Big)^{\!-b}\,,
\end{equation}
where $E(z)=H(z)/H_{0}$, $\Omega_{DE0}=1-\Omega_{m0}$ and the dimensionless constant $A$ is given by
\begin{equation}
A = -\frac{3^{\frac{1}{b}}}{\xi\tilde{\gamma}^{\frac{1}{b}}}
G_{\!N}^{\frac{1}{b}}H_{0}^{\frac{2}{b}}\frac{r_{\Sigma}a_{0}}{\sqrt{G_{\!N}}}\,.
\end{equation}
Obviously the parameter $A$ quantifies the deviation from the concordance $\Lambda$CDM model.
Also, prior to the present epoch $z\to 0$, the Hubble expansion of the current model tends to
that of  $\Lambda$CDM.

Our aim is to constrain the free parameters of the model with the aid of the maximum likelihood
analysis. The latter, in the absence of systematic effects and for Gaussian errors, is identical
to minimizing the $\chi^2$ function in terms of the free parameters $\phi^{\nu}$.
In this case the statistical vector is $\phi^{\nu} = (\Omega_{m0},A,h)$, where $h=H_{0}/100$.
In order to minimize $\chi^2$ with respect to $\phi^{\nu}$, we employ the Markov Chain Monte Carlo
(MCMC) algorithm of Ref. \cite{AffineInvMCMC}, implemented within the Python package emcee
\cite{emcee}. The convergence of the algorithm was checked with auto-correlation time considerations
and also with the Gelman-Rubin criterion \cite{GelmanRubinAll}.

As mentioned above, the theoretically favored values of $b$ ($\Lambda_{k} \propto k^{b}$) are
close to its IR value 2, with indicative such values $b=2.13$ and $b=2.08$ corresponding to
galactic and cluster scales respectively \cite{Kofinas:2017gfv}. Here, we utilize $b=2.08$.
However, we have tested that the likelihood analysis provides very similar results for
a whole range of the $b$ values close to the predicted IR value 2.


\subsection{Data and Methodology}

\subsubsection{Hubble parameter data}

We use the most recent $H(z)$ data as collected by \cite{YuRatra2018}. This set contains $N=36$
measurements of the Hubble expansion in the following redshift range $0.07\leq z\leq 2.33$.
Out of these, there are 5 measurements based on BAOs, while for the rest, the Hubble parameter
is measured via the differential age of passive evolving galaxies. Generally, concerning the
$H(z)$ data, there are two possible ways to proceed, namely (a) to use the measures of
$H(z)$ from the differential ages of passively evolving galaxies (the so called Cosmic
Chronometer data - hereafter CC) and (b) to utilize the full data-set which consists
$H(z)$ measurements from both CC and BAOs observations. In our work we decide to follow
the second avenue. However, we also explore the first avenue, namely to use $H(z)$ data from
CC only and to include a recent BAO dataset to our likelihood analysis (see Appendix for the
relevant discussion). Here, the corresponding $\chi^2_{H}$ function reads
\begin{equation}
\label{chisq:H}
\chi^{2}_{H}\left(\phi^{\nu}\right)=\pmb{\cal H}\,
{\bf C}_{H,\text{cov}}^{-1}\,\pmb{\cal H}^{T}\,,
\end{equation}
where $\pmb{\mathcal{H}}=\{H_{1}-H_{0}E(z_{1},\phi^{\nu})\,,\,...\,,\,
H_{N}-H_{0}E(z_{N},\phi^{\nu})\}$ and $H_{i}$ are the observed Hubble rates at redshifts $z_{i}$
($i=1,...,N$). Notice, that the matrix ${\bf C}$ will denote everywhere the corresponding covariance
matrix. For more details regarding the statistical analysis and the corresponding covariances we
refer the work of \cite{AA} and references therein.

However, we checked that our results are in agreement within $1\sigma$ with the relevant
results from the (a) option along with BAOs data, where we used data and method presented in
\cite{Mehra2015}. This point and the relevant results are discussed further at the next section.

\subsubsection{Standard Candles}

Concerning the standard candles we include in our analysis the binned Pantheon set of Scolnic et al.
\cite{Scolnic:2017caz} and the binned sample of Quasi-Stellar Objects (QSOs) \cite{RisalitiLusso:2015}.
We would like to stress that the combination of the Pantheon SnIa data with those of QSOs opens a
new window in observational cosmology, namely we trace the Hubble relation in the redshift range
$0.07< z <6$, where the dark matter and the dark energy dominate the cosmic expansion.
Notice that the chi-square function of the standard candles is given by
\begin{equation}
\chi^{2}_{\text{s}}\left(\phi^{\nu}_{\text{s}}\right)=\pmb{\mu}_{\text{s}}\,
{\bf C}_{\text{s},\text{cov}}^{-1}\,\pmb{\mu}_{\text{s}}^{T}\,,
\end{equation}
where
$\pmb{\mu}_{\text{s}}=\{\mu_{1}-\mu_{\text{th}}(z_{1},\phi^{\nu})\,,\,...\,,\,
\mu_{N}-\mu_{\text{th}}(z_{N},\phi^{\nu})\}$ and the subscript $\text{s}$ denotes SnIa
and QSOs. For the supernovae data the covariance matrix is not diagonal, while we have
$\mu_{i} = \mu_{B,i}-\mathcal{M}$, where $\mu_{B,i}$ is the approximate distance modulus at
redshift $z_{i}$ and $\mathcal{M}$ is treated as a universal free parameter \cite{Scolnic:2017caz}.
It is easy to observe that $\mathcal{M}$ and $h$ parameters are fundamentally degenerate in
the context of the binned Pantheon data set, so we cannot extract any information regarding
$H_{0}$ from SnIa data alone. In the case of QSOs, $\mu_{i}$ is the observed distant modulus
at redshift $z_{i}$ and the covariance matrix is diagonal. For both SnIa and QSOs, the theoretical
form of the distance modulus reads
\begin{equation}
\mu_{\text{th}} = 5\log\left(\frac{D_{L}}{\text{Mpc}}\right) + 25\,,
\end{equation}
where $D_L = c(1+z)H_0^{-1}\int_{0}^{z}E^{-1}(x,\phi^{\nu})dx$ is
the luminosity distance.

\subsubsection{CMB shift parameters}

Traditionally, every cosmological model that is to be considered as an alternative of
the $\Lambda$CDM needs to be confronted also with CMB shift parameter data. The AS
Eq.~(\ref{Hubble_rate_dan}) is valid for small redshifts since the included anti-gravity
effects emerge only in the presence of structures. However, it is significant for
Eq.~(\ref{Hubble_rate_dan}) to be able to provide a smooth matching with a matter dominated
early universe. Indeed, at large enough redshifts ($z\gg 1$), this equation is approximated
by an effective $\Lambda$CDM model
$H^{2}\approx \Omega_{m0}H_{0}^{2}z^{3}+\Lambda_{\text{eff}}/3$,
where the value of the effective
cosmological constant is $\Lambda_{\text{eff}}=3H_{0}^{2}\big(\Omega_{DE0}^{-1/b}+A\big)^{-b}$.
The successful values of the parameters, to be presented later from the fittings, imply
that $\Lambda_{\text{eff}}$ is of order $H_{0}^{2}$, therefore this cosmological constant term
becomes negligible as $z$ increases. As a result, although we will not enter in this work
into a discussion of the perturbations of the considered model, we expect that these will
remain very close to the perturbations of the concordance $\Lambda$CDM model at higher redshifts
and only present modifications at recent $z$'s. Since measurements of the growth rate
$f\sigma_{8}$ exist for relative small $z<2$, perturbations of the AS model could be calculated
and compared to the $\Lambda$CDM results; however, the background cosmological
Eq.~(\ref{Hubble_rate_dan}) is not sufficient for this, but the full gravity equations
should be derived considering the same AS anti-gravity effects.
Now, an accurate and deep geometrical probe of dark energy is the angular scale of the
sound horizon at the last scattering surface as encoded in the location of the first peak
of the CMB temperature spectrum. This geometrical probe is described by the following set
of CMB shift parameters $(l_{a},\mathcal{R})$
\begin{equation}
l_a = \pi \frac{r(z_{*})}{r_s(z_{*})}
\end{equation}
\begin{equation}
\mathcal{R} = \sqrt{\Omega_{m0}}H_{0}D_{A}(z_{*})c^{-1}\,.
\end{equation}
The quantity $D_{A}$ is the angular diameter distance, namely $D_A = D_L(1+z)^{-2}$, and
$r_s$ is the co-moving sound horizon which is given by
\begin{equation}
r_{s}=\int_{0}^{t}\frac{c_s(t')dt'}{a(t')}=\frac{c}{H_{0}} \int_{0}^{a}\frac{c_s(a')da'}{E(a')a'^2}\,,
\end{equation}
where the sound velocity is $c_s(a)=1/\sqrt{3(1+R_{b}a)}$ with
$R_{b}=31500\Omega_{b0}h^2(T_{\text{CMB}}/2.7K)^{-4}$ and $T_{\text{CMB}}=2.7255K$ \cite{Fixsen:2009ug}.
Also, the redshift of the recombination epoch $z_{*}$ is given from the fitting formula
of \cite{HuSugiyama1996} as
\begin{equation}
\label{HUU}
z_{*}= 1048\left[1+0.00124(\Omega_{b0}h^2)^{-0.738}\right]\times
\left[1+g_1\left(\Omega_{m0} h^2\right)^{g_2}\right]\,,
\end{equation}
where the quantities $g_1,g_2$ are
\begin{equation}
g_1 = \frac{0.0783(\Omega_{b0}h^2)^{-0.238}}{1+39.5\left(\Omega_{b0}h^2\right)^{0.763}}\,\,\,\,\,,
\,\,\,\,\, g_2=\frac{0.560}{1+21.1\left(\Omega_{b0}h^2\right)^{1.81}}
\end{equation}
and $\Omega_{b0}$ is the baryon density parameter at the present time.

We also add to the Hubble rate \eqref{Hubble_rate_dan} the standard radiation term.
The radiation density is taken equal to $\Omega_{r0} = a_{eq}\Omega_{m0}$, where
$a_{eq} = \left(1 + 2.5\Omega_{m0}h^210^4\left(\frac{T_{CMB}}{2.7K}\right)^{-4}\right)^{-1}$
\cite{Wang:2015tua}.

The chi-square estimator now reads
\begin{equation}
\chi^2_{\text{CMB}}=\left(\Delta l_a, \Delta \mathcal{R},
\Delta \Omega_{*} \right)\,{\bf C}_{\text{CMB,cov}}^{-1}\,\left(\Delta l_a, \Delta \mathcal{R},
\Delta \Omega_{*} \right)^{T}\,,
\end{equation}
where $\Delta l_a = l_a - 301.77$, $\Delta \mathcal{R} = \mathcal{R} - 1.7482$,
$\Delta \Omega_{*} = \Omega_{b0} - 0.02226$. The corresponding uncertainties are
$\sigma_{l}=0.090$, $\sigma_{\mathcal{R}} = 0.0048$, $\sigma_{\Omega_{*}}=0.00016$, while
the covariance matrix is $C_{ij} = \sigma_{ij}c_{ij}$, where $\sigma_{ij}$ is
the uncertainty and $c_{ij}$ the elements of the normalized covariance matrix
taken from \cite{Wang:2015tua}.

It is well known that the measured CMB shift parameters are somewhat model dependent,
but mainly for cosmological models that include massive neutrinos or those with a strongly
varying equation of state parameter (which is not our case). For a detailed discussion the
reader may find more information in \cite{Elga2007}.

\subsubsection{Joint analysis and model selection}

In order to perform a joint statistical analysis of $K$ cosmological probes (in our case $K=4$),
we need to use the total likelihood function
\begin{equation}
\mathcal{L}_{\text{tot}}(\phi^{\omega}) = \prod_{k=1}^{K} \exp(-\chi^2_{k})\,.
\end{equation}
This implies that the corresponding  $\chi^2_{\text{tot}}$ expression is written as
\begin{equation}
\chi_{\text{tot}}^2 = \sum_{k=1}^{K}\chi^2_{k}\,,
\end{equation}
where the total statistical vector has dimension $\omega$, which is the sum of the $\nu$
parameters of the model at hand plus the number $\nu_{\text{hyp}}$ of hyper-parameters of the data
sets used, i.e. $\omega = \nu + \nu_{\text{hyp}}$. It is important to note that from a
statistical point of view, the hyper-parameters quantifying uncertainties in the data set and
the free parameters of the cosmological model are equivalent. Moreover, we introduce in our
analysis the so called information criteria  (IC) in order to test the statistical
performance of the models themselves. In particular,
we use the Akaike Information Criterion (AIC) \cite{Akaike1974},
the Bayesian Information Criterion (BIC) \cite{Schwarz1978} and the
Deviance Information Criterion (DIC) \cite{Spiegelhalter2002}.
These criteria quantify the goodness of fit which is affected by
the number of free parameters.
From the theoretical viewpoint, AIC is an asymptotically unbiased estimator of the
Kullback-Leibler information which measures the loss of information during the fit.
The BIC criterion comes from Bayesian foundations, namely as an asymptotic approximation
of the marginal probability (i.e. Bayesian evidence) in the large sample limit (for discussion
see \cite{Burnham2011}). Lastly, the DIC criterion is constructed by combining the above
two approaches \cite{Liddle:2007fy}.

The corrected AIC estimator for small data sets ($N_{\rm tot}/\omega<40$) is given by \cite{Ann2002}
\begin{equation}
\text{AIC}=-2\ln(\mathcal{L}_{\text{max}})+2\omega+
\frac{2\omega(\omega+1)}{N_{\rm tot}-\omega-1}\,,
\end{equation}
where $\mathcal{L}_{\text{max}}$ is the maximum likelihood of the data set(s) under
consideration and $N_{\rm tot}$ is the total number of data. Naturally, for large
$N_{\rm tot}$ this expression reduces to the standard form of the AIC criterion, namely
$\text{AIC} = -2\ln(\mathcal{L}_{\text{max}})+2\omega$.

The Bayesian Information Criterion reads
\begin{equation}
\text{BIC} = -2\ln(\mathcal{L}_{\text{max}})+\omega\,{\rm log}(N_{\text{tot}})\,.
\end{equation}
The AIC and BIC criteria employ only the likelihood value at maximum. In principle, due to
the Bayesian nature of our analysis, the accuracy of $\mathcal{L}_{\text{max}}$ is reduced,
meaning that the AIC and BIC values are meant to be used just for illustrative purposes.
In practice, however, by using long chains we obtain $\mathcal{L}_{\text{max}}$ values with enough
accuracy to use them in order to calculate AIC and BIC. We have explicitly checked that the mean
value of the likelihood is approximately equal to the maximum of the likelihood within $1\sigma$ area.

The Deviance Information Criterion is defined as
\begin{equation}
{\rm DIC} = D(\overline{\phi^\omega}) + 2C_{B}\,.
\end{equation}
The DIC is given in terms of the so called Bayesian complexity $C_{B}$ which measures
the power of data to constrain the free parameters compared to the predictivity of the model
which is given by the prior. Specifically,
$C_{B} = \overline{D(\phi^\omega)} - D(\overline{\phi^\omega})$,
where the overline denotes the usual mean value.
The quantity $D(\phi^\omega)$ is the Bayesian Deviation, where
in our case it reduces to $D(\phi^\omega) = -2\ln(\mathcal{L(\phi^\omega)})$.

An important tool for model selection is the relative difference of the Information
Criterion (IC) value for a number of models,
$\Delta \text{IC}_{\text{model}}=\text{IC}_{\text{model}}-\text{IC}_{\text{min}}$,
where the $\text{IC}_{\text{min}}$ is the minimum $\text{IC}$ of each model under consideration.
If a given model has $\Delta\text{IC}\leq 2$, it is deemed statistically compatible with the
``best'' model, while the condition $2<\Delta\text{IC}<6$ indicates a middle tension between
the two models, and subsequently the condition $\Delta\text{IC}\geq 10$ suggests a strong
tension \cite{KassRaftery1995}.

\section{Observational constraints}

We use the above described data sets and analysis to constrain the free parameters of the
Asymptotically Safe Swiss cheese cosmological model presented in Sec. \ref{SCAS}. For completeness
we also provide the constrains of the flat $\Lambda$CDM model, and using the AIC, BIC and DIC tests,
we compare its statistical performance with that of AS model. Consequently, using the extracted
values of the free parameters, we reconstruct the effective DE equation of state (EoS) of the AS
cosmology and we derive the current value of the deceleration parameter and the transition redshift.


\begin{figure}[ht]
\includegraphics[width=0.7\textwidth]{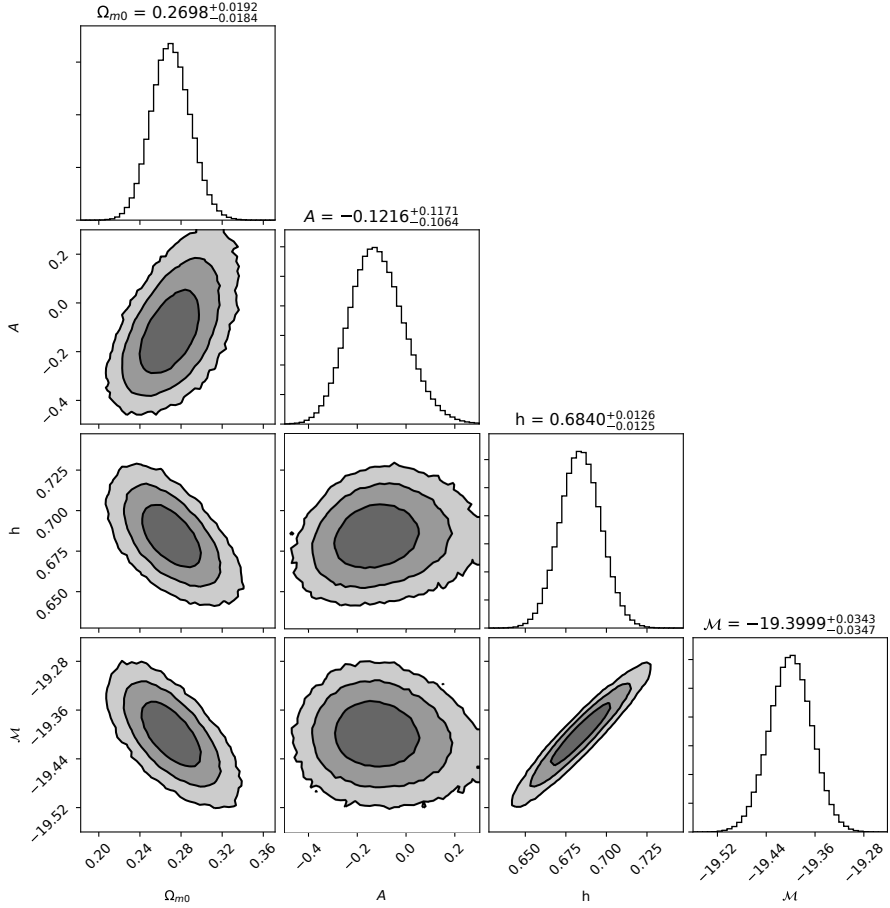}
\caption{The $1\sigma$, $2\sigma$ and $3\sigma$ of the AS cosmological model
for various parameter combinations using
the parameter space $(\Omega_{m0},A,h,\mathcal{M})$. Also, we provide the mean
values of the parameters located in the $1\sigma$ area of the MCMC chain.
Here we combine $H(z)$/Pantheon/QSOs data.}
\label{fig:astroph_probes:Kofinas}
\end{figure}

Combining the direct measurements of the Hubble parameter and the observed Hubble relation given
by Pantheon and the QSOs data ($H(z)$/Pantheon/QSOs), we obtain the following results:
$(\Omega_{m0}, A, h, \mathcal{M}) = \big( 0.270^{+0.019}_{-0.018}\,,\, -0.122
_{-0.106}^{+0.117} \,,\, 0.684^{+0.013}_{-0.012}\,,\,-19.400_{-0.035}^{+0.034} \big)$ with
$\chi^{2}_{\rm min}/dof=84.463/97$.
Concerning the $\Lambda$CDM model
we find $(\Omega_{m0}, h, \mathcal{M}) =
\big( 0.281_{-0.015}^{+0.016},\, 0.686\pm 0.013\,,\,-19.403\pm 0.035 \big)$ with
$\chi^{2}_{\rm min}/dof=85.700/98$.
Inserting the CMB shift parameters data in the analysis, the overall likelihood function peaks at
$(\Omega_{m0}, A, h,\Omega_{b0}h^{2}, \mathcal{M}) = \big( 0.303\pm 0.001\,,\,0.065_{-0.077}^{+0.079} \,,\,0.685
\pm 0.009\,,\,0.0223 \pm 0.0002\,,\,-19.404 \pm 0.020 \big)$
and
$(\Omega_{m0}, h,\Omega_{b0}h^{2}, \mathcal{M}) =
\big( 0.307_{-0.007}^{+0.008}\,,\,0.679
\pm 0.006\,,\,0.0223 \pm 0.0001\,,\,-19.415\pm 0.015\big)$
for AS and $\Lambda$CDM models, respectively, with $\chi^{2}_{\rm min}/dof=89.774/99$
and $\chi^{2}_{\rm min}/dof=90.340/100$.
Evidently, taking into account the $\Delta \text{IC}$s values from Table II, we observe
that while AIC and DIC criteria plead for statistical equivalence between $\Lambda$CDM and AS
cosmology (that is $\Delta$IC$<2$), the BIC criterion supports that there is a mild tension
between the aforementioned models. The apparent discrepancy could be explained by the fact
that the BIC criterion is an asymptotic criterion, taken as $N \rightarrow \infty$. In our
case, the dataset length is not a parameter that must enter the model selection because we
used data compactified with different ways (i.e. binning). Consequently, we chose to consider
the  DIC criterion as the most reliable estimation of the relative probability for each model
to be closer to the reality, based on the fact that it uses all the accessible information for
the likelihood and not just its value at maximum and also bearing less suppositions. With both
DIC and AIC criteria we find $\Delta$IC$<2$ which implies that the AS and the $\Lambda$CDM
models are statistically equivalent. Notice that in Table I we provide a compact presentation
of the current observational constraints, in Table II the relevant values of the information
criteria, while in Fig. 1 and 2 we plot the $1\sigma$, $2\sigma$ and $3\sigma$ confidence
contours in various planes for the AS cosmological model.

We would like to remind the reader that in order to obtain the aforementioned constraints
we imposed $b=2.08$, however, we also checked various other values of $b$ in a range close
and above the theoretically limiting IR value $b=2$. We found that the observational
constraints of the AS model remain practically the same.

It is worth mentioning that our $H(z)$/Pantheon/QSOs/CMB$_{\rm shift}$ best fit values are
in agreement with those of Planck 2018 \cite{Aghanim:2018eyx} (see also \cite{Ade:2015xua}).
Regarding the well known Hubble constant problem, namely the observed Hubble constant
$H_{0}=73.48 \pm 1.66$ Km/s/Mpc found by \cite{RiessHo} is in $\sim 3.7\sigma$ tension with
that of Planck $H_{0} = 67.36 \pm 0.54$ Km/s/Mpc \cite{Aghanim:2018eyx}, we find that the
$H_{0}$ value extracted from the AS model is closer to the latter case. In order to appreciate
the impact of the $H_{0}$ measurement \cite{RiessHo} in constraining the AS model, we check
that the observational constraints (including $H_{0}$) of the combined
$H_{0}$/$H(z)$/SnIa/QSOs/CMB$_{\rm shift}$ statistical analysis are compatible within $1\sigma$
with those of $H(z)$/SnIa/QSOs/CMB$_{\rm shift}$. In particular in the case of
$H_{0}$/$H(z)$/SNIa/QSOs/CMB$_{\rm shift}$ we find
$(\Omega_{m0}, A, h,\Omega_{b0}h^{2}, \mathcal{M}) = \big( 0.291\pm 0.008\,,\,0.145_{-0.071}^{+0.072} \,,\,0.6975
\pm 0.0084\,,\,0.0223 \pm 0.0002\,,\,-19.379 \pm 0.018 \big)$.

Furthermore, we calculate the acceleration parameter and the transition redshift. In order
to calculate the latter quantities, we utilize the best fit parameters provided by the
combination of the cosmological probes (see Table I).
In Fig. 3 we show the evolution of the reconstructed EoS $w_{\rm DE}$ for $H(z)$/Pantheon/QSOs
(upper panel) and $H(z)$/Pantheon/QSOs/CMB$_{\rm shift}$ (lower panel) respectively.
First of all, we observe that the $\Lambda$CDM value $w_{\Lambda}(z)=-1$ is inside the
$1\sigma$ area in the solution space of the AS model.
In the case of $H(z)$/Pantheon/QSOs, the today's value of the acceleration parameter is
$q_0 = -0.595 \pm 0.029$ and the EoS parameter at the present epoch is $w_{DE0} = -0.927 \pm 0.070$.
In this context, the transition redshift is found to be $z_{t} = 0.737 \pm 0.056$. For the
combination $H(z)$/Pantheon/QSOs/CMB$_{\rm shift}$ we find $q_0 = -0.546 \pm 0.015$,
$w_{DE0} = -1.038 \pm 0.046$ and $z_{t} = 0.669 \pm 0.028$. These kinematic parameters are in
agreement (within $1\sigma$ uncertainties) with those of Haridasu et al. \cite{Haridasu2018}
who found, using a model independent way, $z_t = 0.64_{-0.09}^{+0.18}$ and $q_0 = -0.52 \pm 0.06$.
The latter is another indicator of the robustness of the AS Swiss cheese model.

\begin{figure}[ht]
\includegraphics[width=0.7\textwidth]{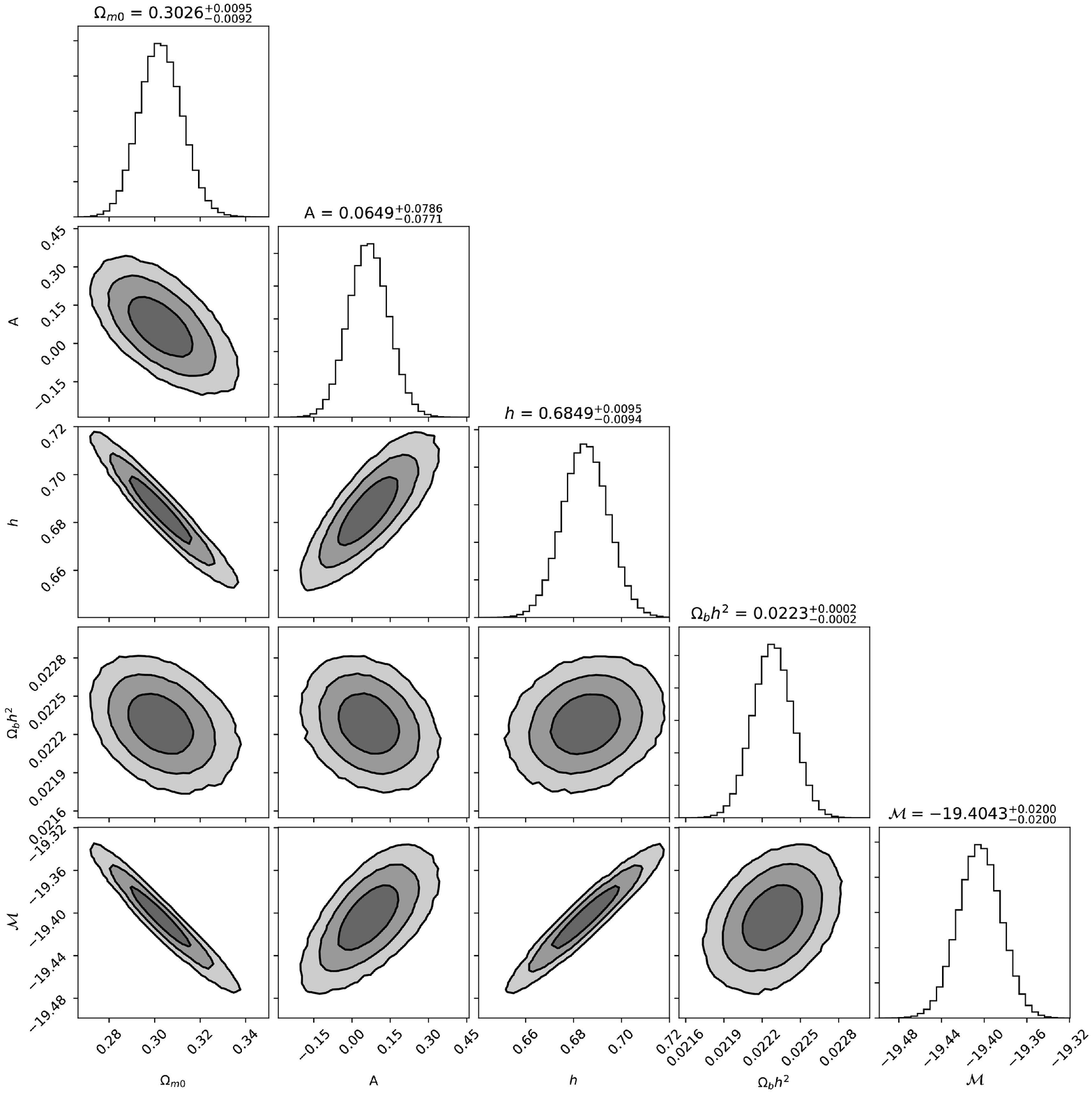}
\caption{The $1\sigma$, $2\sigma$ and $3\sigma$ of the AS cosmological model
for various parameter combinations using
the parameter space $(\Omega_{m0},A,h,\Omega_{b0}h^{2},\mathcal{M})$.
Also, we provide the mean values of the parameters located in the $1\sigma$ area
of the MCMC chain. Here we combine $H(z)$/Pantheon/QSOs/CMB$_{\rm shift}$ data.}
\label{fig:all_probes:Kofinas}
\end{figure}

\begin{figure}[ht]
\includegraphics[width=0.7\textwidth]{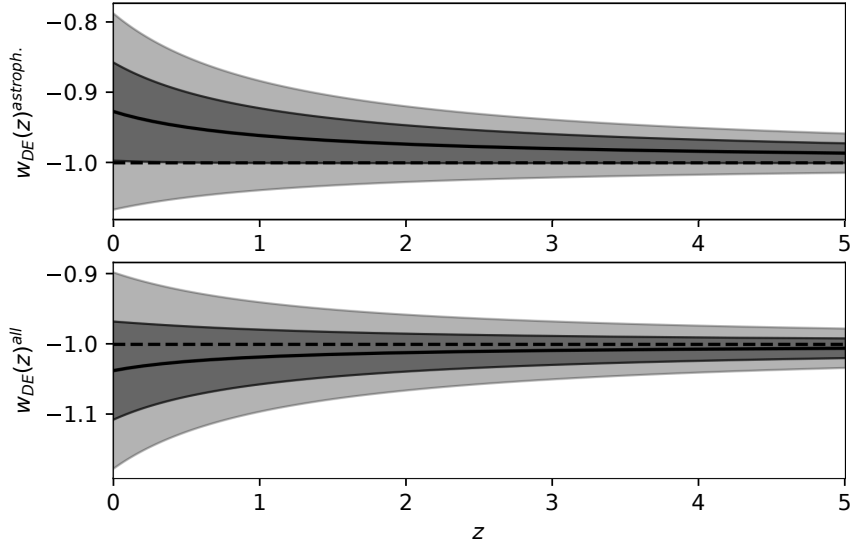}
\caption{The evolution of the effective dark energy equation of
state parameter $w_{DE}(z)$ with the corresponding $1\sigma$ and
$2\sigma$ uncertainties. In the upper panel we use the best fit values from the
combination $H(z)$/Pantheon/QSOs, while in the lower panel we utilize
those of $H(z)$/Pantheon/QSOs/CMB$_{\rm shift}$.
The dashed line corresponds to $\Lambda$CDM value $w=-1$.
Notice that we use the open source Python package uncertainties,
in which the uncertainties are calculated by using first order error propagation theory
from \cite{uncertainties}.}
\label{fig:wz_reconstruction}
\end{figure}

\begin{table*}[!]
\vspace{0.6cm}
\tabcolsep 5.5pt
\vspace{1mm}
\begin{tabular}{ccccccc} \hline \hline
Model & $\Omega_{m0}$ & $A$ & $h$ &$\Omega_{b0}h^{2}$ & $\mathcal{M}$ &
$\chi_{\text{min}}^{2}/\text{dof}$  \vspace{0.05cm}\\ \hline
\hline
 \\
 \multicolumn{6}{c}{\emph{$H(z)$/\text{Pantheon}/\text{QSOs}}}\\ \\
 AS model & $0.270_{-0.018}^{+0.019}$ & $-0.122_{-0.106}^{+0.117} $ & $0.684_{-0.012}^{+0.013} $& - &$-19.400
 _{-0.035}^{+0.034}$ & $ 84.463/96 $ \vspace{0.01cm}\\

$\Lambda$CDM & $0.281^{+0.016}_{-0.015}$ & - & $0.686 \pm 0.013$& - & $-19.403
\pm 0.035$ & $85.700/97$  \vspace{0.45cm}\\ 

 \multicolumn{6}{c}{\emph{$H(z)$/\text{Pantheon}/\text{QSOs}/\text{CMB} \text{shift}}}\\ \\
AS model & $0.303 \pm 0.001 $ & $0.065_{-0.077}^{+0.079} $& $ 0.685 \pm 0.009$ & $0.0223
\pm 0.0002$ & $ -19.404 \pm 0.020$ & $89.774/98$  \vspace{0.01cm}\\ 

$\Lambda$CDM& $0.307^{+0.008}_{-0.007} $ & - & $0.679 \pm 0.006 $& $0.0223 \pm 0.0001$ & $-19.415 \pm 0.015$ & $90.340/99 $ \vspace{0.45cm}\\ 

\hline\hline
\end{tabular}
\caption[]{The observational constraints of the cosmological parameters and the
corresponding $\chi^{2}_{\rm min}/\text{dof}$ for the AS cosmology and $\Lambda$CDM, where dof
are the reduced degrees of freedom, that is $N-\omega$, with $N$ the total number of data
points and $\omega$ the length of the statistical vector.
\label{tab:Results1}}
\end{table*}

\begin{table*}[!]
\vspace{0.5cm}
\tabcolsep 5.5pt
\vspace{1mm}
\begin{tabular}{ccccccc} \hline \hline
Model & AIC & $\Delta$AIC & BIC &$\Delta$BIC & DIC & $\Delta$DIC
 \vspace{0.05cm}\\ \hline
\hline
 \\
 \multicolumn{7}{c}{\emph{$H(z)$/\text{Pantheon}/\text{QSOs}}}\\ \\
 AS model &  92.889 & 0.937 & 102.884 & 3.369 & 92.409 & 0.738  \vspace{0.01cm}\\

$\Lambda$CDM & 91.952 & 0 & 99.515 & 0 & 91.671 & 0 \vspace{0.45cm}\\ 

 \multicolumn{7}{c}{\emph{$H(z)$/\text{Pantheon}/\text{QSOs}/\text{CMB} \text{shift}}}\\ \\
AS model & 100.419 & 1.653 & 112.948 & 4.068 & 99.568 & 1.33  \vspace{0.01cm}\\ 

$\Lambda$CDM & 98.766 & 0  & 108.879 & 0 & 98.238 & 0\vspace{0.45cm}\\ 

\hline\hline
\end{tabular}
\caption[]{The information criteria values AIC, BIC, DIC for the AS cosmology and
$\Lambda$CDM along with the corresponding differences
$\Delta\text{IC} \equiv \text{IC} - \text{IC}_{\text{min}}$.
\label{tab:Results2}}
\end{table*}

\section{Conclusions}
\label{Conclusions}

We study a novel cosmological model introduced in \cite{Kofinas:2017gfv}, which seems to
solve naturally the dark energy issue and its associated cosmic coincidence problem of recent
acceleration without the need to introduce extra scales or fine-tuning. The solution is based
on the existence of anti-gravity sources generated from infrared quantum gravity modifications
at intermediate astrophysical scales (galaxies or clusters of galaxies) according to the
Asymptotic Safety (AS) program. A Swiss cheese metric has been used to extract the cosmological
evolution as the result of the total repulsive effects inside the structures.

Here, we have used the most recent observational data sets, namely $H(z)$, SnIa (Pantheon), QSOs,
BAOs and CMB shift parameters from Planck in order to constraint the free parameters of the AS
cosmological model. We found that the AS cosmological model is very efficient and in excellent
agreement with observations. The derived energy density of matter from the AS model is compatible
with the corresponding quantity derived imposing the concordance cosmology from the CMB angular power
spectrum. Also, the AS model supports a smaller value of Hubble constant than the value obtained
from Cepheids. Specifically, the best fit value for $H_0$ in the AS model is closer to the Planck value.

We also reconstruct the equation of state parameter as a function of the redshift using the derived
values of the free parameters. The today's value of the equation of state parameter is close
to $w=-1$. The transition redshift and the current value of the deceleration parameter
derived here are compatible with the analogous values derived without any assumptions for the
underlying cosmology in the literature.

By comparing the concordance $\Lambda$CDM model and the considered AS model in terms of the
fitting properties using a variety of information criteria, we conclude that the AS model
is statistically equivalent with that of $\Lambda$CDM. To our view, this is an
important result because the AS model, unlike the majority of the cosmological models,
does not include new fields in nature, hence it can be seen as a viable alternative cosmological
scenario towards explaining the accelerated expansion of the universe.
Finally, despite the fact that the AS scenario is statistically equivalent with
that of $\Lambda$CDM at the expansion level, the two models are expected to have differences
at the perturbation level.

\begin{acknowledgments}

FA wishes to thank Angelos Berketis for carefully reading this manuscript and commenting.
SB acknowledges  support  by  the  Research Center for Astronomy of the Academy of Athens in the
context of the program ``Testing general relativity on cosmological scale'' (ref. number 200/872).
GK and VZ acknowledge the support of Orau Small Grants of Nazarbayev University
(faculty-development competitive research grants program for 2019-2021, pure ID 15854058).
VZ acknowledges the hospitality of Nazarbayev University.

\end{acknowledgments}

\section*{Appendix}

Here we combine BAOs in a joint analysis
with the other probes (Cosmic Chronometers, SnIa,
QSOs, CMB shift parameters) in order to test the performance of our results appeared in section IV.
In this case the total $\chi ^2_{\rm tot}$ is given by
\begin{eqnarray}
\label{chisqTot}
\chi^2_{\rm tot} = \chi^2_{\rm CC}+\chi_{\rm SnIa}^2+\chi^2_{\rm QSO}+\chi^{2}_{\rm CMB}
+ \chi^2_{\rm BAO}\,.
\end{eqnarray}
The quantities $\chi_{\rm SnIa}^2$, $\chi^2_{\rm QSO}$, $\chi^{2}_{\rm CMB}$ are defined
in section IIIA, while $\chi^2_{\rm CC}$ is given by Eq.~(\ref{chisq:H}), where we have
used $H(z)$ measurements from the differential ages of passively evolving galaxies
(Cosmic Chronometers). To this end, $\chi^2_{\rm BAO}$ is written as
\begin{eqnarray}
\chi_{\rm BAO}^2 = {\bf D}{\bf C}_{\text{cov}}^{-1}{\bf D}^{T}\,,
\end{eqnarray}
where ${\bf{D}}=\{\lambda (z_{1},\phi^\omega)-d_{1,\text{obs}}\,,\,...\,,\,
\lambda (z_{N},\phi^\omega)-d_{N,\text{obs}}\}$
and ${\bf C}_{\text{cov}}^{-1}$ is the corresponding covariance matrix.
In the latter expression, $\lambda(z_{i},\phi^\omega)$ is defined as follows
\begin{eqnarray}
\label{defininglambda}
\lambda(z_{i},\phi^\omega) =
    \begin{cases}
        \,D_{V}\, r_{d,\text{fid}}/r_{d}\,\,, \  1 \leq i \leq 4\\
        \,D_A/r_d\,\,, \ 5 \leq i \leq 9 \,\,\,\,\,\,\,\,\,\,\,\,\,\,\,\,\,\,\,\,,\\
        \,D_M/r_d\,\,, \ i = 10
    \end{cases}
\end{eqnarray}
where
\begin{subequations}
    \begin{equation}
        D_M = \frac{D_{L}(z,\phi^\omega)}{1+z}
    \end{equation}
    \begin{equation}
        D_A = \frac{D_{L}(z,\phi^\omega)}{(1+z)^2}
    \end{equation}
    \begin{equation}
        D_V = \left[\frac{cD_A^2 \,z (1+z)^2}{H(z,\phi^\omega)}\right]^{1/3}\,.
    \end{equation}
    \end{subequations}

The quantity $r_{d}$ is the standard ruler, a characteristic length scale of the over-densities
in the distribution of matter, therefore Baryonic Acoustic Oscillations are affected by the
standard ruler. For the concordance $\Lambda$CDM model, it is known that
$r_d$ is equal to the co-moving sound horizon $r_{s}$ at the redshift of the baryon drag epoch.
However, for other cosmological models the situation concerning $r_d$ might be different
\cite{Verde:2016ccp}.
Following the notations of \cite{Tutusaus:2017ibk}, it seems natural to treat $r_{d}$ as a free
parameter. In this case the statistical vector takes the following form
$\phi^{\omega}=(\Omega_{m0},A,h,\Omega_{b0}h^2,\mathcal{M},r_d)$. In Table III we list the
BAOs data along with the corresponding references. The final column of Table III includes
$r_{d,\text{fid}}$ which is the sound horizon of the fiducial model ($\Lambda$CDM) used in
order to extract the data. Notice that the corresponding covariances can be found in
\cite{Kazin:2014qga}, \cite{Gil-Marin:2015sqa} and \cite{Gil-Marin:2018cgo}.
The quantity $\lambda$ stands for the observable quantity in each case, defined
by Eq.~(\ref{defininglambda}), while $\sigma_{\lambda}$ is the relative uncertainty.

\begin{table*}[!]
\vspace{1mm}
\tabcolsep 5.5pt
\vspace{1mm}
\begin{tabular}{ccccc}
\hline
 \emph{z} & $\lambda$ & $\sigma_{\lambda}$ & $r_{d,\text{fid}}$ (Mpc) & Ref \\
 \hline
0.122 &  539.0 Mpc      &  17.0      & 147.5  & \cite{Carter:2018vce} \\
0.44  & 1716.40 Mpc    &  83.00      & 148.6  & \cite{Kazin:2014qga}\\
0.60   & 2220.80 Mpc    & 101.00      & 148.6 &  \cite{Kazin:2014qga}\\
0.73  & 2516.00 Mpc      &  86.00      & 148.6 & \cite{Kazin:2014qga} \\
0.32  &    6.67   &   0.13   & 148.11&  \cite{Gil-Marin:2015sqa}\\
0.57  &    9.52   &   0.19   & 148.11&  \cite{Gil-Marin:2015sqa}\\
1.19  &   12.6621 &   0.9876 & 147.78 & \cite{Gil-Marin:2018cgo}\\
1.50   &   12.4349 &   1.0429 & 147.78 &\cite{Gil-Marin:2018cgo} \\
1.83  &   13.1305 &   1.0465 & 147.78 &\cite{Gil-Marin:2018cgo} \\
2.40   &   36.0    &   1.2    &  -   & \cite{Bourboux:2017cbm}\\
\hline
\end{tabular}
\caption{The BAOs dataset.}
\label{tab:BAOS_data}
\end{table*}

\begin{figure}[ht]
\includegraphics[width=0.7\textwidth]{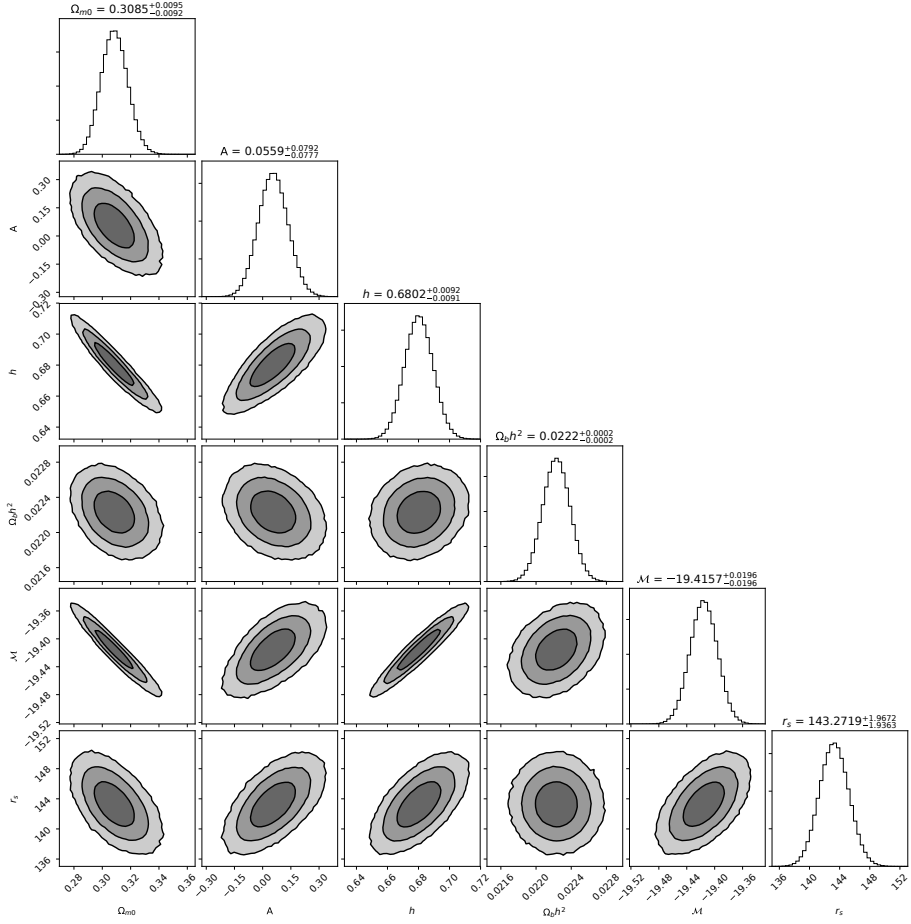}
\caption{The $1\sigma$, $2\sigma$ and $3\sigma$ of the AS cosmological model for various
parameter combinations using the parameter space
$(\Omega_{m0},A,h,\Omega_{b0}h^2,\mathcal{M},r_d)$. Also, we provide the mean values
of the parameters located in the $1\sigma$ area of the MCMC chain. Notice that $r_s$
stands for $r_d$ and is measured in Mpc. Here we combine
CC/Pantheon/QSOs/CMB$_{\rm{shift}}$/BAOs data.}
\label{fig:plusBAOs}
\end{figure}



As it turns out, the overall likelihood function peaks at
$\Omega_{m0}=0.308 \pm 0.09$ , $A=0.0559^{+0.0792}_{-0.0777}$ ,
$h = 0.6802^{+0.0092}_{-0.0091}$ , $\Omega_{b0}h^2 = 0.0222 \pm 0.0002$ ,
$\mathcal{M} = -19.416 \pm 0.010$ and $r_d = 143.272^{+1.967}_{-1.936}$ Mpc,
while the relevant $\chi^2_{\text{min}}/\text{dof}=91.579/102$.
In order to verify the robustness of our results, we explored the effect of using
different data set, i.e. after excluding WiggleZ and data from BOSS DR12 \cite{Gil-Marin:2015sqa},
while adding data from Ref. \cite{Alam:2016hwk} instead. We explicitly verify that
our observational constraints remain unaltered.

In Fig. \ref{fig:plusBAOs} we visualize the solution space by showing the $1\sigma$, $2\sigma$
and $3\sigma$ confidence contours for various parameter pairs. Evidently, there is $1\sigma$
compatibility between the aforementioned results with those of Section IV (see also Table
\ref{tab:Results1}). Finally, it is interesting to mention that our result regarding $r_d$
is in agreement with the independent measurement provided by Verde {\it{et al.}} \cite{Verde:2016ccp},
namely $r_d^{\text{ind}} = 141.0 \pm 5.5$ Mpc.


\begin{thebibliography}{99}

\bibitem{ATLAS:2012oga}
  G.~Aad {\it et al.} [ATLAS Collaboration],
  Science {\bf 338}, 1576 (2012)
  doi:10.1126/science.1232005;
  G.~Aad {\it et al.} [ATLAS Collaboration],
  Phys.\ Lett.\ B {\bf 716}, 1 (2012)
  doi:10.1016/j.physletb.2012.08.020
  [arXiv:1207.7214 [hep-ex]].


\bibitem{Zeldovich:1967gd}
  Y.~B.~Zeldovich,
  JETP Lett.\  {\bf 6}, 316 (1967)
  [Pisma Zh.\ Eksp.\ Teor.\ Fiz.\  {\bf 6}, 883 (1967)].


\bibitem{Riess:1998cb}
A.~G.~Riess {\it et al.} [Supernova Search Team],
Astron.\ J.\  {\bf 116}, 1009 (1998)
doi:10.1086/300499
[astro-ph/9805201].


\bibitem{Perlmutter:1998np}
S.~Perlmutter {\it et al.} [Supernova Cosmology Project Collaboration],
Astrophys.\ J.\  {\bf 517}, 565 (1999)
doi:10.1086/307221
[astro-ph/9812133].


\bibitem{Knop:2003iy}
R.~A.~Knop {\it et al.} [Supernova Cosmology Project Collaboration],
Astrophys.\ J.\  {\bf 598}, 102 (2003)
doi:10.1086/378560
[astro-ph/0309368];
A.~G.~Riess {\it et al.} [Supernova Search Team],
Astrophys.\ J.\  {\bf 607}, 665 (2004)
doi:10.1086/383612
[astro-ph/0402512].


\bibitem{Spergel:2006hy}
D.~N.~Spergel {\it et al.} [WMAP Collaboration],
Astrophys.\ J.\ Suppl.\  {\bf 170}, 377 (2007)
doi:10.1086/513700
[astro-ph/0603449].


\bibitem{Komatsu:2010fb}
E.~Komatsu {\it et al.} [WMAP Collaboration],
Astrophys.\ J.\ Suppl.\  {\bf 192}, 18 (2011)
doi:10.1088/0067-0049/192/2/18
[arXiv:1001.4538 [astro-ph.CO]].


\bibitem{Ade:2013zuv}
P.~A.~R.~Ade {\it et al.} [Planck Collaboration],
Astron.\ Astrophys.\  {\bf 571}, A16 (2014)
doi:10.1051/0004-6361/201321591
[arXiv:1303.5076 [astro-ph.CO]].


\bibitem{Huterer:1998qv}
D.~Huterer and M.~S.~Turner,
Phys.\ Rev.\ D {\bf 60}, 081301 (1999)
doi:10.1103/PhysRevD.60.081301
[astro-ph/9808133].


\bibitem{Aghanim:2018eyx}
  N.~Aghanim {\it et al.} [Planck Collaboration],
  arXiv:1807.06209 [astro-ph.CO].


\bibitem{Ade:2015xua}
  P.~A.~R.~Ade {\it et al.} [Planck Collaboration],
  Astron.\ Astrophys.\  {\bf 594}, A13 (2016)
  doi:10.1051/0004-6361/201525830
  [arXiv:1502.01589 [astro-ph.CO]].


\bibitem{ShapiroSola}
  I.~L.~Shapiro and J.~Sola,
  Phys.\ Lett.\ B {\bf 530}, 10 (2002)
  doi:10.1016/S0370-2693(02)01355-2
  [hep-ph/0104182].


\bibitem{bassola}
  S.~Basilakos, M.~Plionis and J.~Sola,
  Phys.\ Rev.\ D {\bf 80}, 083511 (2009)
  doi:10.1103/PhysRevD.80.083511
  [arXiv:0907.4555 [astro-ph.CO]].


\bibitem{Sool14}
  J.~Sola,
  ``Cosmological constant and vacuum energy: old and new ideas,''
  J.\ Phys.\ Conf.\ Ser.\  {\bf 453}, 012015 (2013)
  doi:10.1088/1742-6596/453/1/012015
  [arXiv:1306.1527 [gr-qc]];
  J.~Sola,
  ``Vacuum energy and cosmological evolution,''
  AIP Conf.\ Proc.\  {\bf 1606}, 19 (2014)
  doi:10.1063/1.4891113
  [arXiv:1402.7049 [gr-qc]];
  J.~Grande, J.~Sola, S.~Basilakos and M.~Plionis,
  JCAP {\bf 1108}, 007 (2011)
  doi:10.1088/1475-7516/2011/08/007
  [arXiv:1103.4632 [astro-ph.CO]].


\bibitem{gomez}
  A.~Gomez-Valent, J.~Sola and S.~Basilakos,
  JCAP {\bf 1501}, 004 (2015)
  doi:10.1088/1475-7516/2015/01/004
  [arXiv:1409.7048 [astro-ph.CO]].


\bibitem{Kofinas:2011pq}
G.~Kofinas and V.~Zarikas,
Eur.\ Phys.\ J.\ C {\bf 73}, no. 4, 2379 (2013)
doi:10.1140/epjc/s10052-013-2379-9
[arXiv:1107.2602 [hep-th]].


\bibitem{Li:2011sd}
  M.~Li, X.~D.~Li, S.~Wang and Y.~Wang,
  Commun.\ Theor.\ Phys.\  {\bf 56}, 525 (2011)
  doi:10.1088/0253-6102/56/3/24
  [arXiv:1103.5870 [astro-ph.CO]].


\bibitem{Kofinas:2017gfv}
  G.~Kofinas and V.~Zarikas,
  Phys.\ Rev.\ D {\bf 97}, no. 12, 123542 (2018)
  doi:10.1103/PhysRevD.97.123542
  [arXiv:1706.08779 [gr-qc]].


\bibitem{ASreviews}
M.~Niedermaier and M.~Reuter,
Living Rev.\ Rel.\  {\bf 9}, 5 (2006);
R.~Percacci,
In {\textit{Oriti, D. (ed.): Approaches to quantum gravity}} 111-128
[arXiv:0709.3851 [hep-th]];
O.~Lauscher and M.~Reuter,
In {\textit{Fauser, B. (ed.) et al.: Quantum gravity}} 293-313 [hep-th/0511260];
M.~Reuter and F.~Saueressig,
New J.\ Phys.\  {\bf 14}, 055022 (2012)
[arXiv:1202.2274 [hep-th]];
A.~Bonanno,
PoS CLAQG {\bf 08}, 008 (2011) [arXiv:0911.2727 [hep-th]];
M.~Niedermaier,
Class.\ Quant.\ Grav.\  {\bf 24}, R171 (2007) [gr-qc/0610018];
R. Percacci, {\textit{``An Introduction to Covariant Quantum Gravity and Asymptotic Safety''}},
Word Scientific, ISBN: 978-981-3207-17-2;
R. Percacci and D. Perini, Phys. Rev. D68, 044018 (2003);
  D.~F.~Litim,
  Phys.\ Rev.\ Lett.\  {\bf 92}, 201301 (2004)
  doi:10.1103/PhysRevLett.92.201301
  [hep-th/0312114];
  K.~Falls, D.~F.~Litim, K.~Nikolakopoulos and C.~Rahmede,
  arXiv:1301.4191 [hep-th];
  K.~Falls, C.~R.~King, D.~F.~Litim, K.~Nikolakopoulos and C.~Rahmede,
  Phys.\ Rev.\ D {\bf 97}, no. 8, 086006 (2018)
  doi:10.1103/PhysRevD.97.086006
  [arXiv:1801.00162 [hep-th]];
  D.~F.~Litim,
  PoS QG {\bf -Ph}, 024 (2007)
  doi:10.22323/1.043.0024
  [arXiv:0810.3675 [hep-th]].


\bibitem{Falls:2014tra}
  K.~Falls, D.~F.~Litim, K.~Nikolakopoulos and C.~Rahmede,
  Phys.\ Rev.\ D {\bf 93}, no. 10, 104022 (2016)
  doi:10.1103/PhysRevD.93.104022
  [arXiv:1410.4815 [hep-th]];
  K.~Falls,
  Phys.\ Rev.\ D {\bf 92}, no. 12, 124057 (2015)
  doi:10.1103/PhysRevD.92.124057
  [arXiv:1501.05331 [hep-th]];
  C.~Ahn, C.~Kim and E.~V.~Linder,
  Phys.\ Lett.\ B {\bf 704}, 10 (2011)
  doi:10.1016/j.physletb.2011.08.075
  [arXiv:1106.1435 [astro-ph.CO]];
  N.~Christiansen, D.~F.~Litim, J.~M.~Pawlowski and M.~Reichert,
  arXiv:1710.04669 [hep-th];
  C.~M.~Nieto, R.~Percacci and V.~Skrinjar,
  Phys.\ Rev.\ D {\bf 96}, no. 10, 106019 (2017)
  doi:10.1103/PhysRevD.96.106019
  [arXiv:1708.09760 [gr-qc]];
  G.~Kofinas and V.~Zarikas,
  Phys.\ Rev.\ D {\bf 94}, no. 10, 103514 (2016)
  doi:10.1103/PhysRevD.94.103514
  [arXiv:1605.02241 [gr-qc]].

\bibitem{Scolnic:2017caz}
  D.~M.~Scolnic {\it et al.},
  Astrophys.\ J.\  {\bf 859} (2018) no.2,  101
  doi:10.3847/1538-4357/aab9bb
  [arXiv:1710.00845 [astro-ph.CO]];
  The numerical data of the full Pantheon SnIa sample are available at
http://dx.doi.org/10.17909/T95Q4X
https://archive.stsci.edu/prepds/ps1cosmo/index.html.



\bibitem{GRB}
  L.~Amati {\it et al.},
  Astron.\ Astrophys.\  {\bf 390}, 81 (2002)
  doi:10.1051/0004-6361:20020722
  [astro-ph/0205230];
  G.~Ghirlanda, G.~Ghisellini and C.~Firmani,
  New J.\ Phys.\  {\bf 8}, 123 (2006)
  doi:10.1088/1367-2630/8/7/123
  [astro-ph/0610248];
  S.~Basilakos and L.~Perivolaropoulos,
  Mon.\ Not.\ Roy.\ Astron.\ Soc.\  {\bf 391}, 411 (2008)
  doi:10.1111/j.1365-2966.2008.13894.x
  [arXiv:0805.0875 [astro-ph]];
  F.~Y.~Wang, Z.~G.~Dai and E.~W.~Liang,
  New Astron.\ Rev.\  {\bf 67}, 1 (2015)
  doi:10.1016/j.newar.2015.03.001
  [arXiv:1504.00735 [astro-ph.HE]].


\bibitem{Plionis:2011jj}
  M.~Plionis, R.~Terlevich, S.~Basilakos, F.~Bresolin, E.~Terlevich, J.~Melnick and R.~Chavez,
  Mon.\ Not.\ Roy.\ Astron.\ Soc.\  {\bf 416}, 2981 (2011)
  doi:10.1111/j.1365-2966.2011.19247.x
  [arXiv:1106.4558 [astro-ph.CO]];
  R.~Chavez, M.~Plionis, S.~Basilakos, R.~Terlevich, E.~Terlevich, J.~Melnick, F.~Bresolin and A.~L.~Gonzalez-Moran,
  Mon.\ Not.\ Roy.\ Astron.\ Soc.\  {\bf 462}, no. 3, 2431 (2016)
  doi:10.1093/mnras/stw1813
  [arXiv:1607.06458 [astro-ph.CO]].


\bibitem{Blake:2011en}
  C.~Blake {\it et al.},
  Mon.\ Not.\ Roy.\ Astron.\ Soc.\  {\bf 418}, 1707 (2011)
  doi:10.1111/j.1365-2966.2011.19592.x
  [arXiv:1108.2635 [astro-ph.CO]];
\bibitem{Alam:2016hwk}
  S.~Alam {\it et al.} [BOSS Collaboration],
  Mon.\ Not.\ Roy.\ Astron.\ Soc.\  {\bf 470}, no. 3, 2617 (2017)
  doi:10.1093/mnras/stx721
  [arXiv:1607.03155 [astro-ph.CO]].


\bibitem{Calabrese:2016bnu}
  E.~Calabrese, N.~Battaglia and D.~N.~Spergel,
  Class.\ Quant.\ Grav.\  {\bf 33}, no. 16, 165004 (2016)
  doi:10.1088/0264-9381/33/16/165004
  [arXiv:1602.03883 [gr-qc]].


\bibitem{Basilakos:2017rgc}
  S.~Basilakos and S.~Nesseris,
  Phys.\ Rev.\ D {\bf 96}, no. 6, 063517 (2017)
  doi:10.1103/PhysRevD.96.063517
  [arXiv:1705.08797 [astro-ph.CO]].


\bibitem{Lavre2018}
  L.~Kazantzidis and L.~Perivolaropoulos,
  Phys.\ Rev.\ D {\bf 97}, no. 10, 103503 (2018)
  doi:10.1103/PhysRevD.97.103503
  [arXiv:1803.01337 [astro-ph.CO]].


\bibitem{Jimenez:2001gg}
  R.~Jimenez and A.~Loeb,
  Astrophys.\ J.\  {\bf 573}, 37 (2002)
  doi:10.1086/340549
  [astro-ph/0106145].


\bibitem{Samushia:2006fx}
  L.~Samushia and B.~Ratra,
  Astrophys.\ J.\  {\bf 650}, L5 (2006)
  doi:10.1086/508662
  [astro-ph/0607301];
  O.~Farooq and B.~Ratra,
  Astrophys.\ J.\  {\bf 766}, L7 (2013)
  doi:10.1088/2041-8205/766/1/L7
  [arXiv:1301.5243 [astro-ph.CO]];
  L.~P.~Chimento, M.~G.~Richarte and I.~E.~Sanchez Garcia,
  Phys.\ Rev.\ D {\bf 88}, 087301 (2013)
  doi:10.1103/PhysRevD.88.087301
  [arXiv:1310.5335 [gr-qc]];
  P.~C.~Ferreira, D.~Pavon and J.~C.~Carvalho,
  Phys.\ Rev.\ D {\bf 88}, 083503 (2013)
  doi:10.1103/PhysRevD.88.083503
  [arXiv:1310.2160 [gr-qc]];
  S.~Capozziello, O.~Farooq, O.~Luongo and B.~Ratra,
  Phys.\ Rev.\ D {\bf 90}, no. 4, 044016 (2014)
  doi:10.1103/PhysRevD.90.044016
  [arXiv:1403.1421 [gr-qc]];
  O.~Akarsu, S.~Kumar, R.~Myrzakulov, M.~Sami and L.~Xu,
  JCAP {\bf 1401}, 022 (2014)
  doi:10.1088/1475-7516/2014/01/022
  [arXiv:1307.4911 [gr-qc]];
  C.~Gruber and O.~Luongo,
  Phys.\ Rev.\ D {\bf 89}, no. 10, 103506 (2014)
  doi:10.1103/PhysRevD.89.103506
  [arXiv:1309.3215 [gr-qc]];
  M.~Forte,
  Gen.\ Rel.\ Grav.\  {\bf 46}, no. 10, 1811 (2014)
  doi:10.1007/s10714-014-1811-2
  [arXiv:1311.3921 [gr-qc]].


\bibitem{Denkiewicz:2014kna}
  T.~Denkiewicz, M.~P.~Dabrowski, C.~J.~A.~P.~Martins and P.~E.~Vielzeuf,
  Phys.\ Rev.\ D {\bf 89}, no. 8, 083514 (2014)
  doi:10.1103/PhysRevD.89.083514
  [arXiv:1402.0520 [astro-ph.CO]];
  R.~G.~Cai, Z.~K.~Guo and T.~Yang,
  Phys.\ Rev.\ D {\bf 93}, no. 4, 043517 (2016)
  doi:10.1103/PhysRevD.93.043517
  [arXiv:1509.06283 [astro-ph.CO]];
  F.~Melia and T.~M.~McClintock,
  Astron.\ J.\  {\bf 150}, 119 (2015)
  doi:10.1088/0004-6256/150/4/119
  [arXiv:1507.08279 [astro-ph.CO]];
  Y.~Chen, S.~Kumar and B.~Ratra,
  Astrophys.\ J.\  {\bf 835}, no. 1, 86 (2017)
  doi:10.3847/1538-4357/835/1/86
  [arXiv:1606.07316 [astro-ph.CO]];
  A.~Mukherjee and N.~Banerjee,
  Phys.\ Rev.\ D {\bf 93}, no. 4, 043002 (2016)
  doi:10.1103/PhysRevD.93.043002
  [arXiv:1601.05172 [gr-qc]];
  R.~C.~Nunes, S.~Pan and E.~N.~Saridakis,
  JCAP {\bf 1608}, no. 08, 011 (2016)
  doi:10.1088/1475-7516/2016/08/011
  [arXiv:1606.04359 [gr-qc]];
  J.~Magana, M.~H.~Amante, M.~A.~Garcia-Aspeitia and V.~Motta,
  doi:10.1093/mnras/sty260
  arXiv:1706.09848 [astro-ph.CO];
  R.~Y.~Guo and X.~Zhang,
  Eur.\ Phys.\ J.\ C {\bf 76}, no. 3, 163 (2016)
  doi:10.1140/epjc/s10052-016-4016-x
  [arXiv:1512.07703 [astro-ph.CO]];
  J.~Sola, A.~Gomez-Valent and J.~de Cruz Perez,
  Astrophys.\ J.\  {\bf 836}, no. 1, 43 (2017)
  doi:10.3847/1538-4357/836/1/43
  [arXiv:1602.02103 [astro-ph.CO]].


\bibitem{Einstein:1946ev}
  A.~Einstein and E.~G.~Strauss,
  Annals Math.\  {\bf 47}, 731 (1946)
  doi:10.2307/1969231.


\bibitem{Israel:1966rt}
  W.~Israel,
  Nuovo Cim.\ B {\bf 44S10}, 1 (1966)
  [Nuovo Cim.\ B {\bf 44}, 1 (1966)]
  Erratum: [Nuovo Cim.\ B {\bf 48}, 463 (1967)]
  doi:10.1007/BF02710419, 10.1007/BF02712210;
G. Darmois, {\em M\'{e}morial des Sciences Math\'{e}matiques\/},
Fascicule XXV (Gauthier-Villars, Paris, 1927), Chap. V.


\bibitem{inho}
  G.~F.~R.~Ellis and W.~Stoeger,
  Class.\ Quant.\ Grav.\  {\bf 4}, 1697 (1987)
  doi:10.1088/0264-9381/4/6/025;
S.~R.~Green and R.~M.~Wald,
Phys.\ Rev.\ D {\bf 83}, 084020 (2011)
doi:10.1103/PhysRevD.83.084020
[arXiv:1011.4920 [gr-qc]];
T.~Buchert, M.~J.~France and F.~Steiner,
Class.\ Quant.\ Grav.\  {\bf 34}, no. 9, 094002 (2017),
doi:10.1088/1361-6382/aa5ce2
[arXiv:1701.03347 [astro-ph.CO]];
V.~Marra, E.~W.~Kolb and S.~Matarrese,
Phys.\ Rev.\ D {\bf 77}, 023003 (2008)
doi:10.1103/PhysRevD.77.023003
[arXiv:0710.5505 [astro-ph]];
S.~Rasanen,
EAS Publ.\ Ser.\  {\bf 36}, 63 (2009)
doi:10.1051/eas/0936008
[arXiv:0811.2364 [astro-ph]].


\bibitem{Reuter:2009kq}
M.~Reuter and H.~Weyer,
 Gen.\ Rel.\ Grav.\  {\bf 41}, 983 (2009) [arXiv:0903.2971 [hep-th]];
 M.~Reuter and H.~Weyer,
Phys.\ Rev.\ D {\bf 79}, 105005 (2009) [arXiv:0801.3287 [hep-th]];
 P.~F.~Machado and R.~Percacci,
 Phys.\ Rev.\ D {\bf 80}, 024020 (2009) [arXiv:0904.2510 [hep-th]];
 E.~Manrique and M.~Reuter,
 PoS CLAQG {\bf 08}, 001 (2011) [arXiv:0905.4220 [hep-th]].


\bibitem{Reuter:2001ag}
  M.~Reuter and F.~Saueressig,
  Phys.\ Rev.\ D {\bf 65}, 065016 (2002)
  doi:10.1103/PhysRevD.65.065016
  [hep-th/0110054];
  M.~Reuter,
  Phys.\ Rev.\ D {\bf 57}, 971 (1998)
  doi:10.1103/PhysRevD.57.971
  [hep-th/9605030];
  A.~Bonanno and M.~Reuter,
  JCAP {\bf 0708}, 024 (2007)
  doi:10.1088/1475-7516/2007/08/024
  [arXiv:0706.0174 [hep-th]].


\bibitem{Bonanno:2001hi}
  A.~Bonanno and M.~Reuter,
  Phys.\ Lett.\ B {\bf 527}, 9 (2002)
  doi:10.1016/S0370-2693(01)01522-2
  [astro-ph/0106468];
  A.~Bonanno and M.~Reuter,
  Int.\ J.\ Mod.\ Phys.\ D {\bf 13}, 107 (2004)
  doi:10.1142/S0218271804003809
  [astro-ph/0210472];
  E.~Bentivegna, A.~Bonanno and M.~Reuter,
  JCAP {\bf 0401}, 001 (2004)
  doi:10.1088/1475-7516/2004/01/001
  [astro-ph/0303150];
  I.~Donkin and J.~M.~Pawlowski,
  arXiv:1203.4207 [hep-th];
  D.~Litim and A.~Satz,
  arXiv:1205.4218 [hep-th].


\bibitem{Koch:2014cqa}
  B.~Koch and F.~Saueressig,
  Int.\ J.\ Mod.\ Phys.\ A {\bf 29}, no. 8, 1430011 (2014)
  doi:10.1142/S0217751X14300117
  [arXiv:1401.4452 [hep-th]].


\bibitem{Bonanno:2001xi}
  A.~Bonanno and M.~Reuter,
  Phys.\ Rev.\ D {\bf 65}, 043508 (2002)
  doi:10.1103/PhysRevD.65.043508
  [hep-th/0106133];
  K.~Falls, D.~F.~Litim and A.~Raghuraman,
  Int.\ J.\ Mod.\ Phys.\ A {\bf 27}, 1250019 (2012)
  doi:10.1142/S0217751X12500194
  [arXiv:1002.0260 [hep-th]];
  K.~Falls and D.~F.~Litim,
  Phys.\ Rev.\ D {\bf 89}, 084002 (2014)
  doi:10.1103/PhysRevD.89.084002
  [arXiv:1212.1821 [gr-qc]];
  D.~F.~Litim and K.~Nikolakopoulos,
  JHEP {\bf 1404}, 021 (2014)
  doi:10.1007/JHEP04(2014)021
  [arXiv:1308.5630 [hep-th]];
  D.~F.~Litim,
  Phil.\ Trans.\ Roy.\ Soc.\ Lond.\ A {\bf 369}, 2759 (2011)
  doi:10.1098/rsta.2011.0103
  [arXiv:1102.4624 [hep-th]].


\bibitem{Kofinas:2015sna}
G.~Kofinas and V.~Zarikas,
JCAP {\bf 1510}, no. 10, 069 (2015)
doi:10.1088/1475-7516/2015/10/069;
  C.~Bambi, D.~Malafarina and L.~Modesto,
  Phys.\ Rev.\ D {\bf 88}, 044009 (2013)
  doi:10.1103/PhysRevD.88.044009
  [arXiv:1305.4790 [gr-qc]];
D.~Malafarina,
 Universe {\bf 3}, no. 2, 48 (2017)
 doi:10.3390/universe3020048
 [arXiv:1703.04138 [gr-qc]].


\bibitem{AffineInvMCMC}
J. Goodman and J. Weare, Comm. App. Math. and Comp. Sci. {\bf 5}, 65 (2010).


\bibitem{emcee}
  D.~Foreman-Mackey, D.~W.~Hogg, D.~Lang and J.~Goodman,
  Publ.\ Astron.\ Soc.\ Pac.\  {\bf 125}, 306 (2013)
  doi:10.1086/670067
  [arXiv:1202.3665 [astro-ph.IM]].


\bibitem{GelmanRubinAll}
  A.~Gelman and D.~B.~Rubin,
  Statist.\ Sci.\  {\bf 7}, 457 (1992)
  doi:10.1214/ss/1177011136;
S. P. Brooks and A. Gelman, Jour. Comput. and Graph. Statistics {\bf 7}, 4 (1997).


\bibitem{YuRatra2018}
  H.~Yu, B.~Ratra and F.~Y.~Wang,
  Astrophys.\ J.\  {\bf 856}, no. 1, 3 (2018)
  doi:10.3847/1538-4357/aab0a2
  [arXiv:1711.03437 [astro-ph.CO]].


\bibitem{AA}
  F.~K.~Anagnostopoulos and S.~Basilakos,
  Phys.\ Rev.\ D {\bf 97}, no. 6, 063503 (2018)
  doi:10.1103/PhysRevD.97.063503
  [arXiv:1709.02356 [astro-ph.CO]];
  S.~Basilakos, S.~Nesseris, F.~K.~Anagnostopoulos and E.~N.~Saridakis,
  arXiv:1803.09278 [astro-ph.CO].



\bibitem{Mehra2015}
  A.~Mehrabi, S.~Basilakos and F.~Pace,
  Mon.\ Not.\ Roy.\ Astron.\ Soc.\  {\bf 452}, no. 3, 2930 (2015)
  doi:10.1093/mnras/stv1478
  [arXiv:1504.01262 [astro-ph.CO]].

\bibitem{RisalitiLusso:2015}
  G.~Risaliti and E.~Lusso,
  Astrophys.\ J.\  {\bf 815}, 33 (2015)
  doi:10.1088/0004-637X/815/1/33
  [arXiv:1505.07118 [astro-ph.CO]];
  C.~Roberts, K.~Horne, A.~O.~Hodson and A.~D.~Leggat,
  arXiv:1711.10369 [astro-ph.CO].

\bibitem{Fixsen:2009ug}
  D.~J.~Fixsen,
  Astrophys.\ J.\  {\bf 707}, 916 (2009)
  doi:10.1088/0004-637X/707/2/916
  [arXiv:0911.1955 [astro-ph.CO]].

\bibitem{HuSugiyama1996}
  W.~Hu and N.~Sugiyama,
  Astrophys.\ J.\  {\bf 471}, 542 (1996)
  doi:10.1086/177989
  [astro-ph/9510117].


\bibitem{Wang:2015tua}
  Y.~Wang and M.~Dai,
  Phys.\ Rev.\ D {\bf 94}, no. 8, 083521 (2016)
  doi:10.1103/PhysRevD.94.083521
  [arXiv:1509.02198 [astro-ph.CO]].


\bibitem{Elga2007}
 O.~Elgaroy and T.~Multamaki,
  Astron.\ Astrophys.\  {\bf 471}, 65 (2007)
  doi:10.1051/0004-6361:20077292
  [astro-ph/0702343 [ASTRO-PH]];
  P.~S.~Corasaniti and A.~Melchiorri,
  Phys.\ Rev.\ D {\bf 77}, 103507 (2008)
  doi:10.1103/PhysRevD.77.103507
  [arXiv:0711.4119 [astro-ph]].


\bibitem{Akaike1974}
H. Akaike,
IEEE Transactions on Automatic Control \textbf{19}, 716 (1974).


\bibitem{Schwarz1978}
G. Schwarz, Ann. Statist., \textbf{6}, 2 (1978), 461-464.


\bibitem{Spiegelhalter2002}
Spiegelhalter, David J., et al. Jour. of the R. Stat. Soc., \textbf{64} 4 (2002): 583-639.


\bibitem{Burnham2011}
Burnham, Kenneth P., \emph{et al}. Behavioral Ecology and Sociobiology, \textbf{65}, 1, (2011):2335.


\bibitem{Liddle:2007fy}
  A.~R.~Liddle,
  Mon.\ Not.\ Roy.\ Astron.\ Soc.\  {\bf 377}, L74 (2007)
  doi:10.1111/j.1745-3933.2007.00306.x
  [astro-ph/0701113].


\bibitem{Ann2002}
K. Anderson, {\it Model selection and multimodel inference:
a practical information-theoretic approach}, 2nd edn. Springer, New York (2002);
K. Anderson, Sociological Methods and Research {\bf 33}, 261 (2004).


\bibitem{KassRaftery1995}
Kass, Robert E., and Adrian E. Raftery, Journal of the american statistical association
90.430 (1995): 773-795.


\bibitem{RiessHo}
A. G. Riess, \emph{et al.},
Astrophys.\ J.\  {\bf 855}, 18 (2018)
doi:10.3847/1538-4357/aaadb7
[arXiv:1801.01120 [astro-ph.SR]].


\bibitem{Haridasu2018}
  B.~S.~Haridasu, V.~V.~Lukovic, M.~Moresco and N.~Vittorio,
  arXiv:1805.03595 [astro-ph.CO].


\bibitem{uncertainties}
Uncertainties: a Python package for calculations with uncertainties,
Eric O. LEBIGOT, https://pythonhosted.org/uncertainties/.


\bibitem{Verde:2016ccp}
  L.~Verde, J.~L.~Bernal, A.~F.~Heavens and R.~Jimenez,
  Mon.\ Not.\ Roy.\ Astron.\ Soc.\  {\bf 467}, no. 1, 731 (2017)
  doi:10.1093/mnras/stx116
  [arXiv:1607.05297 [astro-ph.CO]].


\bibitem{Tutusaus:2017ibk}
  I.~Tutusaus, B.~Lamine, A.~Dupays and A.~Blanchard,
  Astron.\ Astrophys.\  {\bf 602}, A73 (2017)
  doi:10.1051/0004-6361/201630289
  [arXiv:1706.05036 [astro-ph.CO]].


\bibitem{Carter:2018vce}
  P.~Carter, F.~Beutler, W.~J.~Percival, C.~Blake, J.~Koda and A.~J.~Ross,
  doi:10.1093/mnras/sty2405
  [arXiv:1803.01746 [astro-ph.CO]].


\bibitem{Kazin:2014qga}
  E.~A.~Kazin {\it et al.},
  Mon.\ Not.\ Roy.\ Astron.\ Soc.\  {\bf 441}, no. 4, 3524 (2014)
  doi:10.1093/mnras/stu778
  [arXiv:1401.0358 [astro-ph.CO]].


\bibitem{Gil-Marin:2015sqa}
  H.~Gil-Marin {\it et al.},
  Mon.\ Not.\ Roy.\ Astron.\ Soc.\  {\bf 460}, no. 4, 4188 (2016)
  doi:10.1093/mnras/stw1096
  [arXiv:1509.06386 [astro-ph.CO]].


\bibitem{Gil-Marin:2018cgo}
  H.~Gil-Marin {\it et al.},
  Mon.\ Not.\ Roy.\ Astron.\ Soc.\  {\bf 477}, no. 2, 1604 (2018)
  doi:10.1093/mnras/sty453
  [arXiv:1801.02689 [astro-ph.CO]].


\bibitem{Bourboux:2017cbm}
  H.~du Mas des Bourboux {\it et al.},
  Astron.\ Astrophys.\  {\bf 608}, A130 (2017)
  doi:10.1051/0004-6361/201731731
  [arXiv:1708.02225 [astro-ph.CO]].










\end{thebibliography}
\end{document}